\documentclass[11pt]{article}

\usepackage{epsfig,graphics,graphicx,amsmath,amssymb,amsfonts}
\usepackage{array,pdftexcmds}
\usepackage[update,prepend]{epstopdf}
\usepackage{overpic}

\usepackage{commath}
\usepackage{subfig}
\usepackage{tabularx}
\usepackage{enumitem}
\usepackage{color}

\newcommand{\bm}[1]{\mbox{\boldmath{$#1$}}}

\def\e{{\bm e}}
\def\f{{\bm f}}
\def\br{{\bm r}}
\def\bv{{\bm v}}
\def\0{{\bm 0}}

\def\cl {\nonumber \\}
\def\el {\nonumber }

\begin{document}
	
	\title{A microscopic approach to study the onset of a highly infectious disease spreading}
	
	\author{Krithika Rathinakumar$^*$ and Annalisa Quaini$^*$ \\
		\footnotesize{$^*$Department of Mathematics, University of Houston, 3551 Cullen Blvd, Houston TX 77204}\\
		\footnotesize{krithika.trk@gmail.com; quaini@math.uh.edu}
	}
	
	\maketitle

%%%%%%%%%%%%%%%%%%%%%%%%%%%%%%%%%%%%%%%%%%%%%%%%%%%%%%%%%%%%%%%
%%%%%%%%%%%%%%%%%%%%%%%%%%%%%%%%%%%%%%%%%%%%%%%%%%%%%%%%%%%%%%% 

\noindent{\bf Abstract}

We combine a pedestrian dynamics model with a contact tracing method to simulate 
the initial spreading of a highly infectious airborne disease in a confined environment. 
We focus on a medium size population (up to 1000 people) with a small number of infectious people (1 or 2)
and the rest of the people is divided between immune and susceptible.
%Most mathematical epidemiology models predict the basic reproduction number, i.e. 
%the number of infected people from an infectious person, in a large population where all the people are susceptible.
%Such models are not valid when the number of infected individuals is small, and the size of the healthy population is medium.
We adopt a space-continuous model that represents pedestrian dynamics by the forces acting on them, 
i.e.~a microscopic force-based model. 
Once discretized, the model results in a high-dimensional system of second order ordinary differential equations. 
Before adding the contact tracing to the pedestrian dynamics model, we calibrate the model parameters, 
validate the model against empirical data, and show that pedestrian self-organization into lanes
can be captured. 
We consider an explicit approach for contact tracing by introducing a sickness domain
around a sick person. A healthy but susceptible person who remains in the sickness domain 
for a certain amount of time may get infected (with a prescribed probability) and become a so-called 
secondary contact. As a concrete setting to simulate the onset of disease spreading, we consider terminals in two US airports:
Hobby Airport in Houston and the Atlanta International Airport.
We consider different scenarios and we quantify the increase in average number of secondary contacts increases
as a given terminal becomes more densely populated, the percentage of immune people decreases,
the number of primary contacts increases, and areas of high density (such as the boarding buses) are present.

\vskip .2cm
\noindent{\bf Keywords:} Crowd dynamics, contact tracing, disease spreading, complex systems

%%%%%%%%%%%%%%%%%%%%%%%%%%%%%%%%%%%%%%%%%%%%%%%%%%%%%%%%%%%%%%%%
\section{Introduction}
\label{sec:MainIntro}
%Introduction

The focus of this work is to study the initiation of disease spreading in a confined, yet complex environment (e.g. an airport terminal),
and in a medium size population (up to 1000 people) with a small number of infectious people (1 or 2).
Standard epidemic theory models describe the evolution of disease spreading in terms of groups
that are susceptible to, infected with and recovered from a particular disease. 
The SIR (Susceptible Infected Recovered - see, e.g., \cite{Kermack1927, Rohani2000}) and SIS 
(Susceptible Infected Susceptible - see, e.g., \cite{Hethcote1984,Garnett1996}) models are the 
foundation of almost all of mathematical epidemiology. 
To quantify the transmission dynamics through contact with healthy individuals, these models use the basic reproduction number, which
measures the number of infected people (secondary contacts)
from an infectious person (primary contact) in a population where all the people are susceptible.
Most models predicting the severity of an epidemic estimate the basic reproduction number for large population sizes. 
See, e.g., \cite{Browne2015, Klinkenberg2006, Hyman2003} and references therein. Such models are typically not valid 
when the number of infected individuals is small and the size of healthy population is medium. 
Medium sized population are typical of confined environments like airports and hospital waiting rooms,
which are the most likely transmittal places during the initial stage of disease spreading. See, e.g., \cite{measles_terminal1,measles_terminal2}
Understanding contact tracing associated with such environments is of paramount importance for an early suppression of an epidemic. 

%\anna{Can you write a small paragraph on the measles cases in CA and at Hobby?}
%
%\kri{A recent instance in September 2018 where a person with measles spent few hours in the Hobby airport across two days shows an application of contact tracing. The Houston Health department contacted people from both the flights taken by the individual and were able to ensure that the disease did not spread further. Other instances in California, Washington and New York also had measles cases brought through flight travel, eventually resulting in wide spread cases within the states. The model described in this paper would augment contact tracing and provide an estimate on the rate of growth of a disease, helping plan the strategy to contain it.}

We propose a simple approach for contact tracing built on a microscopic pedestrian dynamics model. 
Usually, contact tracing is meant as a disease control strategy in which the people who have been in close 
contact to infectious persons are traced. These traced people are monitored so that if they 
become symptomatic they can be efficiently isolated and the disease transmission can be contained \cite{Eichner2003, Fraser2004}.
The effectiveness of contact tracing as a control strategy for ebola and tuberculosis have been studied
theoretically in, e.g., \cite{Rivers2014,Guzzetta2015}. 
Previous works on contact tracing use a wide range of methodologies from individual based models on specific networks 
to compartmental ordinary differential equations at the population level. See, e.g., \cite{Muller2000, Kiss2005, Klinkenberg2006, Ball2011, Browne2015} and references therein. Many differential equation models incorporate contact tracing implicitly 
(see, e.g. \cite{Mubayi2010,Rivers2014}).
In our simulations, contact tracing is limited to pedestrians
coming in contact during a short time span (up to 1 hour).
The contact tracing method is combined with a pedestrian dynamics model 
to estimate the number of people that could potentially be infected 
by a few sick people around them, thereby providing a tool to study the onset of disease spreading.

A very large variety of models have been developed over the years
to describe the complex dynamical behavior of pedestrian crowds. 
The different mathematical models can be divided into three main categories depending
on the (macroscopic, mesoscopic, microscopic) scale of observation \cite{Bellomo2011}. 
Macroscopic models (see, e.g., \cite{HUGHES2002507,5773492})
are suitable for high density, large-scale systems. Thus, they will not be considered for the proposed work. 
The mesoscale approach \cite{Bellomo2011383,Agnelli2015,Bellomo2012,Bellomo2013_new,Bellomo2017_book,Bellomo2015_new,Bellomo2016_new,Bellomo2017_book,Bellomo2019_new,Bellomo2013,KimQuaini2019} derives a Boltzmann-type evolution equation for the statistical
distribution function of the position and velocity of the pedestrians,
in a framework close to that of the kinetic theory of gases.
Microscopic force-based models, which use Newtonian mechanics to interpret pedestrian movement 
as the physical interaction between the people and the environment, are one of the most popular modeling paradigms of continuous models because they describe the movement of pedestrians well qualitatively. See, e.g., \cite{BS:BS3830360405, Helbing1995, 1367-2630-1-1-313,TurnerPenn,Moussad2755,Chraibi2011425,6248013,6701214} and references therein.
Collective phenomena, like unidirectional or bidirectional flow in a corridor \cite{MA20102101,TAJIMA2002709,5339199}, lane formation \cite{Helbing1995,Helbing2004180,Yu2005}, oscillations at bottlenecks \cite{Helbing1995,Helbing2004180},
the faster-is-slower effect \cite{Lakoba01052005,Parisi2007343}, emergency evacuation from buildings \cite{Helbing2004180,Yu2005,LIU20091921}, are well reproduced. Other advantages of these methods are 
the ease of implementation, and in particular parallel implementation, and the fact that they permit higher resolution for 
geometry and time. 
Because of these advantages, we choose to build our contact tracing method on a 
microscopic force-based model first presented in \cite{Chraibi2010}.
%Both force-based and agent-based models may introduce artifacts due to the force representation of human behavior, leading to unrealistic backward movement or oscillating trajectories. These artifacts can be reduced by incorporating extra rules and/or by elaborate calibrations, at the price of increasing the computational cost.

In most of the references cited above, models have been shown to replicate various cases of
pedestrian movement qualitatively through analysis and/or numerical simulations.
The \emph{quantitative} validation of pedestrian flow models is complicated
by the lack of reliable experimental data. In addition, the few available datasets show
large differences \cite{Schadschneider2011,Seyfried2009,Zhang2011}. 
After a calibration of the model parameters, we validate the model in \cite{Chraibi2010}
by comparing with empirical data from \cite{Zhang2011}. We obtain a good quantitative
agreement between the computed and measured fundamental diagrams for a set of 9
experiments, all involving unidirectional motion in a corridor. 
We run a series of tests to understand self-organized lane formation, 
as this is relevant to the kind of environment we focus on. 
Finally, we mention that adjustment of the parameters and data assimilations have also 
been proposed in \cite{Ward150703,JOHANSSON_HELBING} to make evolutionary models more reliable.

The validated pedestrian dynamics model is then combined with a simple method to trace contact.
One or two sick people, referred to as primary contacts, are introduced in the 
pedestrian population. The people infected by the primary contact are called secondary contacts. 
We consider an explicit approach for contact tracing by introducing a sickness domain
around a primary contact. A healthy but susceptible person who remains in that sickness domain 
for a certain amount of time may get infected (with a prescribed probability) and become a secondary contact. 
As a concrete setting to simulate the onset of disease spreading, we consider terminals in two US airports:
Hobby Airport in Houston and the Atlanta International Airport. 
We consider different scenarios: variable population size, variable percentage of immune (non-susceptible) population, 
boarding bridges or boarding buses.
Through the numerical simulations, we quantify the increase in average number of secondary contacts increases
as a given terminal becomes more densely populated, the percentage of immune people decreases,
the number of primary contacts increases, and areas of high density (such as the boarding buses) are present.

%\kri{ This paper explores an area not commonly ventured into in the past. The domains used are from real physical structures (airports), and the people were simulated to mimic human behavior. Building on this, a simple logical contact tracing model has been used to trace the spread of an infectious disease. For each scenario, we have computed the basic reproduction number, which helps understand the intensity of the spread. Our results quantitatively show that, as expected, the disease spreads to more people if more infected people are present, or if the population in the airport is larger. It also shows that areas of high density such as buses are especially hot-spots for disease spread. }

%Outline of this paper
The outline of this paper is as follows. 
In {Sec.~\ref{sec:Model}}, we introduce the force-based microscopic model for pedestrian dynamics we will build upon, we 
describe the numerical method, calibrate the model parameters and validate the model against experimental data.
In {Sec.~\ref{sec:Laneformation}}, we show that the model is capable of reproducing self-organization of pedestrians.
In {Sec.~\ref{sec:CTIntro}}, we present our contact tracing method and combine it with the 
pedestrian dynamics model to simulate the initial spreading of a highly infectious airborne disease.
We report the average number of people infected by one or two sick people in terminals of Hobby Airport 
and Atlanta International Airport for different scenarios.
Conclusions are drawn in Sec.~\ref{sec:Conclusion}.

%%%%%%%%%%%%%%%%%%%%%%%%%%%%%%%%%%%%%%%%%%%%%%%%%%%%%%%%%%%%%%%%
% Section 2 = Microscopic Model Definition

\section{A microscopic force-based model for pedestrian dyanmics}
\label{sec:Model}
 
We briefly present the model we focus on, which was introduced in \cite{Chraibi2010}.
Let us consider a group of $N$ pedestrians in a bounded geometry $\Omega$. 
Each pedestrian is modeled as a circular disk with a given radius.
The dynamics of each pedestrian over a time interval of interest $(0, T]$
is modeled using Newton's second law, i.e.~for pedestrian $i$ with mass $m_i$
and center of mass at $\br_i$ the law of motion is:
\begin{align}\label{eq:law_of_motion}
m_i \ddot{\br}_i = \f_i, \quad i = 1, \dots, N,
\end{align}
where $\f_i$ represents the total forces acting on the pedestrian. 
Source term $\f_i$ includes the force driving the pedestrian towards
their target and the repulsive forces acting on the pedestrian from
other pedestrians, walls, and other obstacles. %, to prevent collisions and overlapping.
Finding an appropriate description of $\f_i$ to obtain realistic pedestrian motion
is not a trivial task.

Given $h > 0$, the boundary $\partial \Omega$ is represented as a set of $N_b$ points: 
$\mathcal{B} = \{ \br_k \in \partial \Omega\}_{k = 0}^{N_b}$ with $|| \br_{k+1} - \br_{k} || = h$, for $k = 0, \dots, N_b - 1$.
The set of boundary points acting on pedestrian $i$ at time $t \in (0, T]$ is:
\begin{align}
\mathcal{B}_i = \{ j \in \mathbb{N}, j \leq N_b:   \br_j \in \mathcal{B} ~ \text{and}~|| \br_j - \br_i || \leq r_w \}, \el
\end{align}
where $\left \| . \right \|$ denotes the Euclidean norm in $\mathbb{R}^2$ and $r_w$ is a cutoff radius for pedestrian-wall interaction. 
The set of all pedestrians that influence the motion of pedestrian $i$ at a certain time is:
\begin{align}
\mathcal{P}_i = \{ j \in \mathbb{N}, j \leq N ~:~|| \br_j - \br_i || \leq r_p \}, \el
\end{align}
where $r_p$ is a cutoff radius for pedestrian-pedestrian interaction.
We assume that the total forces $\f_i$ consist of three contributions:
\begin{align}\label{eq:forces}
\f_i = \f_i^{tar} + \sum_{j \in \mathcal{P}_i} \f_{ij}^{ped} + \sum_{j \in \mathcal{B}_i} \f_{ij}^{bou}, \quad i = 1, \dots, N,
\end{align}
where $\f_i^{tar}$ is the force driving pedestrian $i$ to their target, $\f_{ij}^{ped}$ is the repulsive
force pedestrian $j$ exerts on pedestrian $i$, and $\f_{ij}^{bou}$ is the repulsive force
due to the domain boundary. 
Repulsive forces $\f_{ij}^{ped}$ and $\f_{ij}^{bou}$ model the pedestrians' attempt to avoid collisions and contact 
%with other pedestrians and boundary (i.e., wall and objects) 
by changing their direction. 

The driving force models the intention of a pedestrian to reach a
destination with a certain desired speed $\overline{v}_i$: 
\begin{align}\label{eq:targetforce}
\f_i^{tar} = m_i \frac{\overline{v}_i \e_i - \bv_i}{\tau} , 
\end{align} 
where $\e_i$ is the unit vector directed from pedestrian $i$ to their target,
$\bv_i = \dot{\br}_i$ is the velocity of pedestrian $i$, and $\tau$ is a time constant. 
To represent more involved paths, we generate a sequence of ``checkpoints'' along the path
and for each checkpoint $j$ we specify a radius $r_j$. Checkpoint $j$ is 
considered to be reached when the pedestrian is within a distance $r_j$ of it.
Once a path is assigned to a pedestrian, the target is the first checkpoint along the path and
when the first checkpoint is reached the target is updated to the second checkpoint, and so on.  

In order to define repulsive force $\f_{ij}^{ped} $ in \eqref{eq:forces}, we need to introduce some notation.
The vector connecting pedestrian $i$ with pedestrian $j$, directed
from $i$ to $j$, and the corresponding unit vector are denoted by:
\begin{align}
\br_{ij} = \br_j - \br_i, \quad \e_{ij} = \frac{\br_{ij}}{||\br_{ij}||}. \el
\end{align}
We assume that pedestrian $i$ has an effective diameter $d_i$ that depends linearly on their velocity:
\begin{align}\label{eq:diam}
d_i(\bv_{i}) = d_i^0 + \tau_d ||\bv_{i}||,
\end{align}
$d_i^0$ being their diameter at rest and $\tau_d$ being a proportionality parameter. Eq.~\eqref{eq:diam}
accounts for the fact that a faster pedestrian has an effective larger diameter since he/she will
keep obstacles and other pedestrians at a larger distance. The effective distance between 
pedestrians $i$ and $j$ is then: 
\begin{align}\label{eq:dist}
d_{ij} = \| \br_{ij} \|-\frac{1}{2}(d_i (\bv_i)+  d_j(\bv_j)).
\end{align}
We can now write the repulsive force as:
\begin{align}\label{eq:rep_force}
\f_{ij}^{ped}= -m_i k_{ij} \frac{(\mu \overline{v}_i + v_{ij})^2}{d_{ij}} \e_{ij},
\end{align} 
where $\mu$ is a parameter used to tune the strength of the force, 
$v_{ij}$ is the component of the velocity of $i$ relative to $j$ in the direction of $\e_{ij}$:
\begin{align}\label{eq:vij}
v_{ij} = \frac{1}{2} \left[ (\bv_i - \bv_j) \cdot \e_{ij} + \left|  (\bv_i - \bv_j) \cdot \e_{ij} \right| \right] =
\left\{
\begin{array}{ll}
(\bv_i - \bv_j) \cdot \e_{ij} & \text{if}~ (\bv_i - \bv_j) \cdot \e_{ij} > 0,   \\
0 & \text{otherwise,}
\end{array}
\right.
\end{align}
and $k_{ij}$ is a coefficient that reduces the action-field of the repulsive force 
to the angle of vision of each pedestrian (i.e., $180^{\circ}$):
\begin{align}\label{eq:kij}
k_{ij}= \frac{1}{2}\frac{{\bv_i} \cdot \e_{ij}+ | \bv_i \cdot \e_{ij} |}{\| \bv_i  \|} =
\left\{
\begin{array}{ll}
\frac{ \bv_i \cdot \e_{ij}}{\| \bv_i  \|} & \text{if}~ \bv_i \cdot \e_{ij} > 0~\text{and}~\| \bv_i  \| \neq 0,    \\
0 & \text{otherwise.}
\end{array}
\right. 
\end{align}  

As is intuitive, repulsive force \eqref{eq:rep_force} is directed in the opposite direction
of $\e_{ij}$ and its modulus is inversely proportional to the effective distance between
pedestrians $i$ and pedestrian $j$. 
Moreover, the strength of the repulsive force $\f_{ij}^{ped}$ depends on the angle between $\bv_i$ and $\e_{ij}$.
In fact, the coefficient $k_{ij}$ takes its maximum value (i.e., 1) when pedestrian $i$ is 
moving in the same direction as $\e_{ij}$ and it takes its minimum value (i.e., 0) when the
angle between $\bv_i$ and $\e_{ij}$ is bigger than $90^{\circ}$. 
Notice that, thanks to the definition of $v_{ij}$, pedestrian $i$ feels the 
repulsive force due to pedestrian $j$ only if they are moving toward each other.
So, e.g., if pedestrian $j$ is close to pedestrian $i$, but faster than and ahead of 
$i$, then $\f_{ij}^{ped} = \0$.
The term $\mu  \overline{v}_i$ at the numerator in eq.~\eqref{eq:rep_force} prevents collisions
when the distance between the two pedestrian is small and the relative speed is low, which would lead 
to an otherwise small repulsive force. %This is motivated by the observation that
%pedestrians with a large desired speed $v^0_i$ need stronger repulsive forces to avoid
%collisions. 
The case $\mu = 0$ corresponds to the centrifugal force model introduced in \cite{Yu2005},
which is known to lead to realistic results only if supplemented with a collision detection technique. 
See also \cite{Chraibi2009} for details.

In order to define $\f_{ij}^{bou}$, we note that 
the repulsive force between a pedestrian $i$ and a wall is
zero if $i$ is walking parallel to the wall. In the model though, this is not enough to avoid 
very small repulsive forces when the pedestrians walks almost parallel to
the wall. For this reason, we assume that each pedestrian $i$ feels the repulsive action 
of three points lying on the boundary: the closest boundary point to pedestrian $i$ denoted by $\br_k$, 
and the two neighboring points $\br_{k-1}$ and $\br_{k+1}$ provided that 
$|| \br_k - \br_i || \leq r_w$. If indeed $|| \br_k - \br_i || \leq r_w$, then $\mathcal{B}_i = \{\br_{k-1}, \br_k, \br_{k+1} \}$,
otherwise $\mathcal{B}_i = \emptyset$.
We assume that the repulsive force exerted by boundary point $j \in \mathcal{B}_i$ on pedestrian $i$ is given by:
\begin{align}\label{eq:bou_force}
\f_{ij}^{bou}= -m_i k_{ij} \frac{(\mu_w \overline{v}_i + v^n_{i})^2}{d^{bou}_{ij}} \e_{ij},
\end{align} 
where $k_{ij}$ is defined in~\eqref{eq:kij}, $ v^n_{i}$ is the component of the velocity normal to the boundary and
\begin{align}
d_{ij}^{bou} = \| \br_{ij} \|-\frac{1}{2}d_i (\bv_i). \el
\end{align}

The complete mathematical model is given by (\ref{eq:law_of_motion}), (\ref{eq:forces}), (\ref{eq:targetforce}), (\ref{eq:rep_force}), (\ref{eq:bou_force}).

%%%%%%%%%%%%%%%%%%%%%%%%%%%%%%%%%%%%%%%%%%%%%%%%%%%%%%%%%%%%%%%%
% Section 3 = Numerical Method 

\subsection{Numerical Method}
\label{sec:NumericalMethod}

%Now that the mathematical model for pedestrian motion has been defined, we describe the numerical method that we will implement for the simulations. Model Eqs. \ref{eq:law_of_motion}, \ref{eq:forces}, \ref{eq:targetforce}, \ref{eq:rep_force}, \ref{eq:bou_force} are discretized in time and we start off by calibrating the free parameters $\tau$, $d_i^0$, $\tau_d$,  $r_p$ and $r_w$ of our model to a simple geometry - a corridor. Special focus is on calibrating the interaction constant $\mu$. Once the parameter values are set we move on to validating the model. Two experiments are conducted - one taken from empirical study \cite{Zhang2011} and another from reference \cite{Agnelli2015}. An experiment from \cite{Helbing2002} is also conducted to observe a self organizing behavior of pedestrian motion, i.e. lane formation.  
%
%%%%%%%%%%%%%%%%%%%%%%%%%%%%%%%%%%%%%%%%%%%%%%%%%%%%%%%%%%%%%%%%%%%%%%%%%%%%%%%%%%%
%\subsection{Time discretization}
%\label{subsec:TimeDiscretization}

We introduce the time-discretization step $\Delta t > 0$ and set  $t^n = n \Delta t$, for $n = 1, \ldots, N_t$, with $N_t = T/ \Delta t$.
Moreover, we denote by $y^n$ the approximation of a generic quantity $y$ at the time $t^n$.

Each pedestrian $i$, with $i = 1, \dots, N$, is assigned an initial position $\br_i^0$ and an initial velocity ${v}_i^0$. 
The position at time $t^{n+1}$, with $n \geq 0$ is found with the following centered finite difference 
approximation of eq.~\eqref{eq:law_of_motion}:
\begin{align}\label{eq:discrete_law}
m_i \frac{\br_i^{n+1} - 2 \br_i^n + \br_i^{n-1}}{\Delta t^2} = \f_i^{n}, \quad n = 0,\dots, N_t - 1, i = 1, \dots, N,
\end{align}
where $\f_i^{n}$ is an approximation of $\f_i$ in eq.~\eqref{eq:forces} at time $t^{n}$. 
%\anna{The forces are treated explicitly, right? Does it make sense to make it implicit or semi-implicit?}
Notice that for $n = 0$ in eq.~\eqref{eq:discrete_law} we need $\br_i^{-1}$, which is
computed as follows:
\begin{align}
\br_i^{-1} =  \br_i^{0} - \Delta t \bv_i^0, \quad i = 1, \dots, N. \el
\end{align}
The velocity of each pedestrian at time $t^{n+1}$ is computed by:
\begin{align}\label{eq:velocity_update}
\bv_i^{n+1} =  \frac{\br_i^{n+1} - \br_i^n}{\Delta t}, \quad i = 1, \dots, N.
\end{align}

%\textbf{Code Implementation: }
The results presented in Sec.~{subsec:Calibration}, \ref{subsec:MacroValidation} and 
\ref{sec:Laneformation} have been obtained with an implementation of the above scheme 
in MATLAB \cite{MATLAB}. The simulations rans on a shared 40-core computing server with 512 GB RAM.

%%%%%%%%%%%%%%%%%%%%%%%%%%%%%%%%%%%%%%%%%%%%%%%%%%%%%%%%%%%%%%%%
% Section 4 = Calibration

\subsection{Calibration of the model parameters}
\label{subsec:Calibration}

\begin{figure}
	\centering
	\includegraphics[width=.75\textwidth]{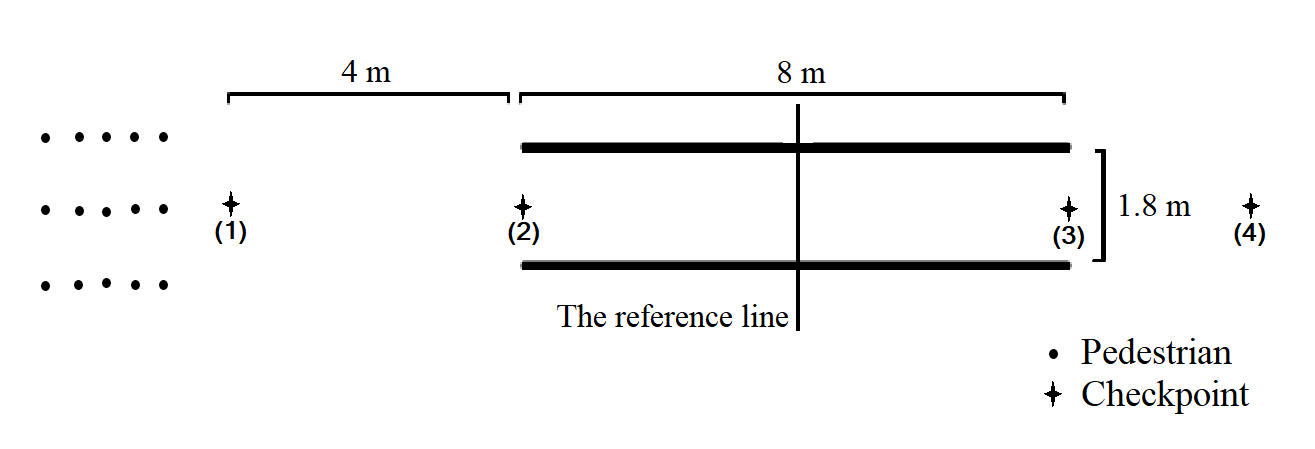}
	\caption{Schematic diagram of the computational domain for the simulations in Sec.~\ref{subsec:Calibration} and \ref{subsec:MacroValidation}.}
	\label{fig:flux_cor}
\end{figure}

The mathematical model described in Sec.~\ref{sec:Model} depends on several parameters. 
To understand the model sensitivity to these parameters parameters, 
we consider the following test. Let $\Omega$ be an $8$ m long and $1.8$ m wide corridor.  
Initially, the pedestrians are placed as shown in Fig.~\ref{fig:flux_cor}: 4 m from the corridor entrance and $1$ m apart from each other. 
People are  initially at rest, i.e. $\bv_i^0 = \mathbf{0}$ for $i = 1, \dots, N$. Every pedestrian is assigned the same path: checkpoint 1 to 4, 
and the radius associated with each of these checkpoints is the corridor width. The desired speed of all pedestrians are Gaussian distributed with mean $1.55$ m/s and standard deviation $0.18$ m/s  ~\cite{Zhang2011}. 

When the simulation is run with the parameters set according to \cite{Chraibi2010} 
(i.e., $\tau = 0.5$ s, $\tau_d = 0.53$ s, $r_p = r_w = 2$ m, $d_i^0 = 0.18$ m, and $\mu = \mu_w = 0.2$), 
some pedestrians attain a speed greater than their desired speed. That case 
makes force \eqref{eq:targetforce} change sign and consequently such pedestrians
move in the direction opposite to their destination. 
Furthermore, pedestrian-pedestrian and pedestrian-wall overlaps occur for 
a large amount of the time interval under consideration. Thus, the parameter values in \cite{Chraibi2010} 
are suitable for unidirectional motion in a narrow corridor. 

The sensitivity analysis for the model parameters $\tau_d$, $\tau$ and $d_i^0$ has been analyzed in detail in \cite{T-Krithika}. 
Based on the results therein, we set $\tau_d = 0.20$ s, $\tau = 0.50$ s, $d_i^0 = 0.18$ m and $\Delta t$ = 0.01 s from now onwards.
Here, we just present the calibration of the cutoff radii $r_p$ and $r_w$ and interaction constant $\mu$ in \eqref{eq:rep_force} and $\mu_w$ in \eqref{eq:bou_force} using the aforementioned test.  

Let us start with $r_p$ and $r_w$. Following  \cite{Chraibi2010}, we set $\mu = \mu_w = 0.2$ for the moment.
We take a group of $N = 12$ people and simulate their passage through the corridor. The reason why we choose 
a small crowd is because in a large group of pedestrians the interaction forces become dominant 
and it is hard to understand the role of other model parameters.
The simulation is run for $T$ = $13$ s, which is the time it takes all the pedestrians to exit the corridor.
We take all the combinations of values for $r_p$ and $r_w$ reported in Table \ref{table:rprw}. 
We consider a simulation unstable if the speed of one pedestrian exceeds their desired speed or overlaps (of people)
and oscillations (of trajectories) occur. From Table \ref{table:rprw}, we see that if either of the radii is small, i.e.~1 m, 
the system is unstable. 
Among the stable combinations, from now on we consider $r_p = r_w = 2$ m because it the cheapest computationally.

\begin{table}[]
	\begin{center}
		\begin{tabular}{|c|c|c|c|}
			\hline
			&$r_p = 1$ m & $r_p = 2$ m & $r_p = 3$ m \\
			\hline
			$r_w = 1$ m & unstable & unstable & unstable \\ 
			\hline
			$r_w = 2$ m & unstable & stable & stable \\ 
			\hline
			$r_w = 3$ m & unstable & stable & stable \\ 
			\hline
		\end{tabular}
	\end{center}	
	\caption{Stability results for different cutoff radii $r_p$ and $r_w$ values.}
	\label{table:rprw}
\end{table}

Next, let us consider the interaction constants in repulsive forces $\mu$ in \eqref{eq:rep_force} and $\mu_{w}$ in \eqref{eq:bou_force}. For simplicity, we consider $\mu = \mu_{w}$. We note that large values of $\mu$ help avoid overlaps between 
pedestrians $i$ and $j$. On the other hand, when $v_{ij}$ is large and $d_{ij}$ is small, a large value of $\mu$ may give rise to a strong repulsive force that compels a pedestrian to deviate more than 90 degrees away from the target direction, leading
oscillations in their trajectory.
Our goal is to find a value for $\mu$ that is large enough to avoid overlaps and small enough to avoid oscillations.
For this purpose, we increase the number of pedestrians to  $N = 36$. For different values of $\mu =$ 0, 0.1, 0.2, 0.3, 0.4, 0.5, 0.6, we run 
$100$ simulations and compute overlap and oscillation quantities defined below.

Following \cite{Chraibi2010}, we define
\begin{align}\label{eq:ov-pro}
	O_v = \frac{1}{n_{ov}} \sum_{t=0}^{t=T} \sum_{i=1}^{i=N} \sum_{j>i}^{j=N} o_{ij}, \quad
    \text{with}~o_{ij}= \frac{A_{ij}}{\min(A_{i},A_{j})} \leq 1, 
\end{align}
where $o_{ij}$ quantifies the ``overlap-strength''and $n_{ov}$ is the cardinality of the set $\{o_{ij} :o_{ij} \neq 0\}$. $A_i$ and $A_j$ 
are the areas of the discs of pedestrians $i$ and $j$, and $A_{ij}$ is their area of intersection. If $n_{ov} =0$, i.e.~no overlap occurs, then $O_v$ is set to $0$. Notice that the maximum value of $O_v$ is 1.

The oscillation-proportion of a simulation is defined as 
\begin{align}\label{eq:os-pro}
	O_s = \frac{1}{n_{os}} \sum^{t=T}_{t=0}\sum_{i=1}^{i=N}S_{i}, \quad \text{with}~S_{i} = \frac{1}{2}(-s_i+\left | s_i \right| ) \quad \text{and}~s_{i} = \frac{v_{i}\cdot ( \overline{v_{i}} \e_i) )}{\overline{v_{i}}^2}, 
\end{align}
where $n_{os}$ is the cardinality of the set $\{S_{i} :S_{i} \neq 0\}$. $S_i$ can be viewed as ``oscillation-strength'' of pedestrian $i$. If $n_{os} =0$, i.e.~no oscillation occurs, then $O_s$ is set to $0.$ Similarly to $O_v$, the maximum value for  $O_s$ is 1.

\begin{figure}
	\centering
	\includegraphics[width=0.75\textheight]{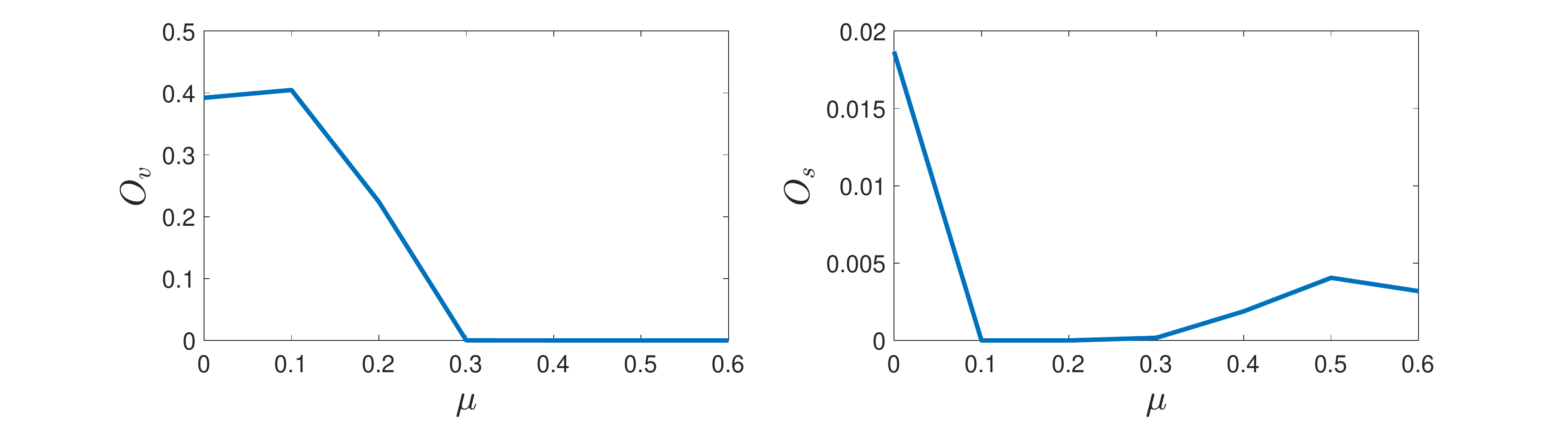}
	\caption{$O_v$  \eqref{eq:ov-pro} and $O_s$  \eqref{eq:os-pro} against the interaction constant $\mu$.}
	\label{fig:osandov}
\end{figure}

Fig.~\ref{fig:osandov} shows the average values of $O_v$ and $O_s$ for $100$ simulations against $\mu$. 
We note that for $\mu = 0, 0.1$ most of the simulation are unstable 
(the speed of some pedestrians exceeded their desired speed). We report the results anyways, although they have little
significance. From Fig.~\ref{fig:osandov}, we see that $O_v$ decreases as $\mu$, as expected. 
Aside from the critical values $\mu = 0, 0.1$, $O_s$ increases as $\mu$ increases, as expected. 
The ideal choice appears to be $\mu = \mu_w =$ 0.3, which yields no overlap and no oscillations. 
However, we consider $\mu = \mu_w =$ 0.2 an acceptable choice too. 
Since pedestrians are modeled as discs with a radius that varies with the pedestrian's speed,
we can assume that a pedestrian does not physically occupy the entire disc. Thus, depending on the test
we allow some amount of overlaps in the system.   

%%%%%%%%%%%%%%%%%%%%%%%%%%%%%%%%%%%%%%%%%%%%%%%%%%%%%%%%%%%%%%%%
% Section 5 = Validation

\subsection{Validation Against Experimental Data}
\label{subsec:MacroValidation}

We \emph{quantitatively} validate described in model Sec.~\ref{sec:Model}, with the parameters set as explained 
in Sec.~\ref{subsec:Calibration}, by comparing with empirical data from \cite{Zhang2011}.
The experiment geometry is the same as in Fig.~\ref{fig:flux_cor}. A perpendicular line passing through the corridor midline is taken as a reference line. Every pedestrian is assigned the same path: checkpoints 1 to 4. Checkpoint 2 and 4 each have a radius of $1.8$ m (width of the corridor), denoted by $b_{cor}$. Checkpoint 1 has radius $b_{ent}$ and checkpoint 3 has radius $b_{exit}$. $N$ pedestrians are placed at a distance of $4.5$ m from the corridor entrance. The values of $b_{ent}$, $b_{exit}$, and $N$ varies for the different experiments. See Table \ref{table:macro}. 

\begin{table}[h]
	\begin{center}
		\begin{tabular}{|c|c|c|c|}
			\hline
			Experiment Index & $N$ & $b_{ent}$ $[m]$ & $b_{exit}$ $[m]$  \\
			\hline 
			1 & 60 & 0.50 & 1.80 \\
			\hline
			2 & 66 & 0.60 & 1.80 \\ 
			\hline
			3 & 111 & 0.70 & 1.80 \\
			\hline
			4 & 121 & 1.00 & 1.80 \\
			\hline
			5 & 175 & 1.45 & 1.80 \\
			\hline
			6 & 220 & 1.80 & 1.80 \\
			\hline
			7 & 170 & 1.80 & 1.20 \\
			\hline
			8 & 160 & 1.80 & 0.95 \\
			\hline
			9 & 148 & 1.80 & 0.70 \\
			\hline		  
		\end{tabular}
		\caption{Number of people $N$, entrance width $b_{ent}$, and exit width $b_{exit}$ for the 
		9 experiments under consideration.}	
		\label{table:macro} 
	\end{center} 
\end{table}

Over a time interval of length $\delta t= 10$ s, the macroscopic quantities flux $J_{\delta t}$, average velocity $v_{\delta t}$ and density $\rho_{\delta t}$ are calculated as follows:
\begin{align}\label{eq:avg-flux-vel-density}
	J_{\delta t} =  \frac{N_{\delta t}}{t_{N_{\delta t}}}, \quad v_{\delta t} =  \frac{1} {N_{\delta t}}\sum_{i=1}^{i=N_{\delta t}} v_i , \quad 
	\rho_{\delta t} =  \frac{J_{\delta t} }{v_{\delta t} b_{cor}}.
\end{align}
where $N_{\delta t}$ is the total number of people who crossed the reference line during $\delta t$, $t_{N_{\delta t}}$ is the time taken by these $N_{\delta t}$ pedestrians to cross the reference line, and $v_i$ is the velocity of the $i^{th}$ pedestrian at the time they cross the reference line. 
For every experiment, we will compare the computed and measured quantities defined in \eqref{eq:avg-flux-vel-density}. 

\begin{figure}
	\centering
	\includegraphics[width=0.5\textwidth]{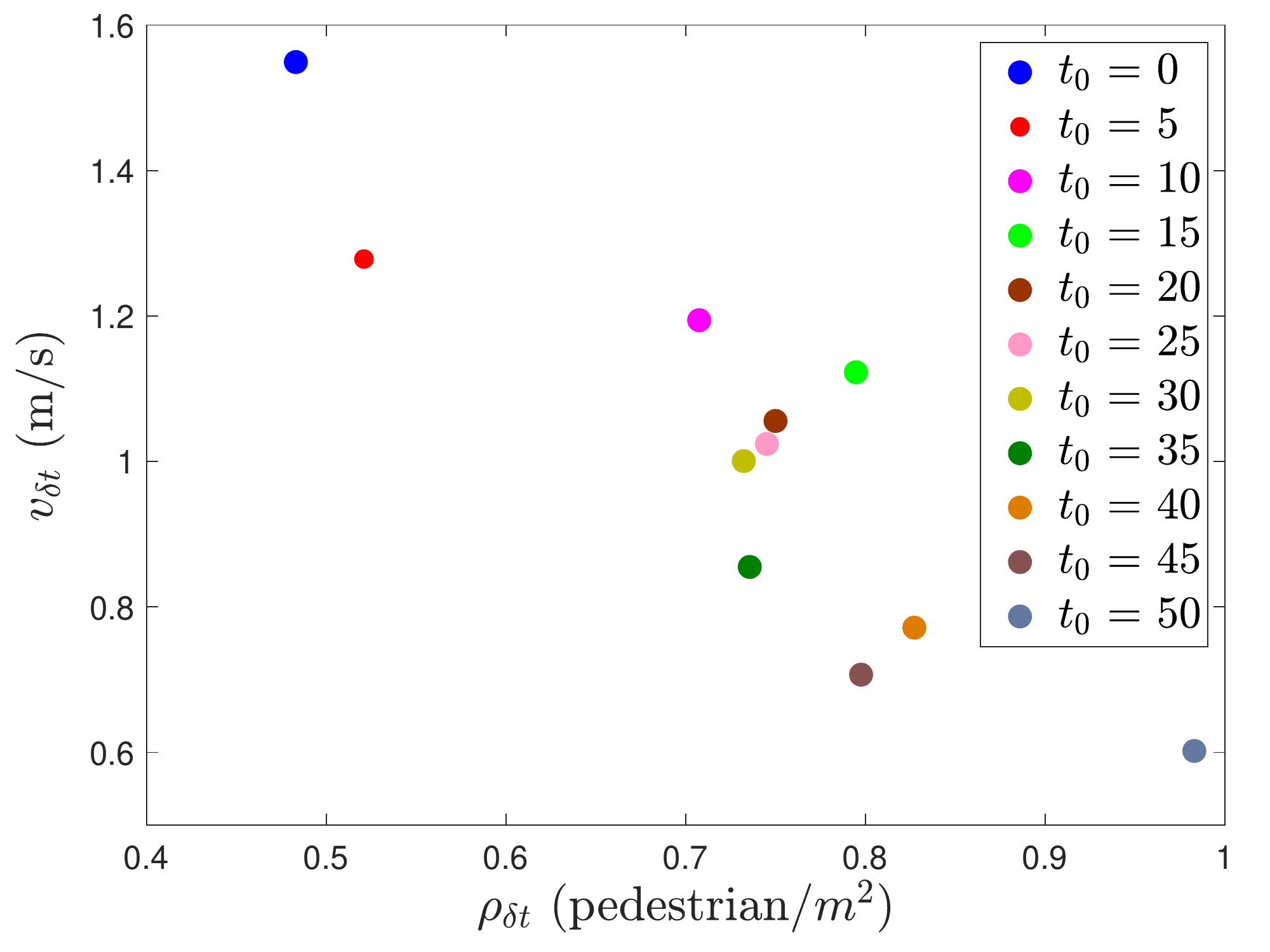}
	\caption{Experiment 1 in Table \ref{table:macro}: average velocity $v_{\delta t}$ \eqref{eq:avg-flux-vel-density} plotted against 
	density $\rho_{\delta t}$ \eqref{eq:avg-flux-vel-density}
	computed using different values of the initial time for interval $\delta t$.} 
	\label{fig:50_den}
\end{figure}

Let $\delta t = [t_0$ ,  $t_0 + \delta t]$. To decide which $t_0$ to pick, we consider experiment 1 in Table \ref{table:macro}. 
Fig. \ref{fig:50_den} shows the computed average velocity plotted against density
for different $t_0$. We see that when $t_0$ is in $[10,30] $ s, the density values are close 
to each other. Thus, we set $t_0 = 10$ s for experiment 1. The values of $t_0$ for all the other
experiments in Table \ref{table:macro} are set in a similar manner. Then, 
the densities and average velocities are averaged over $6$ runs per experiment. 

\begin{figure}[]
	\begin{center}	
		\begin{tabular}{c c}
			\subfloat[]{\includegraphics[width=.4\textheight]{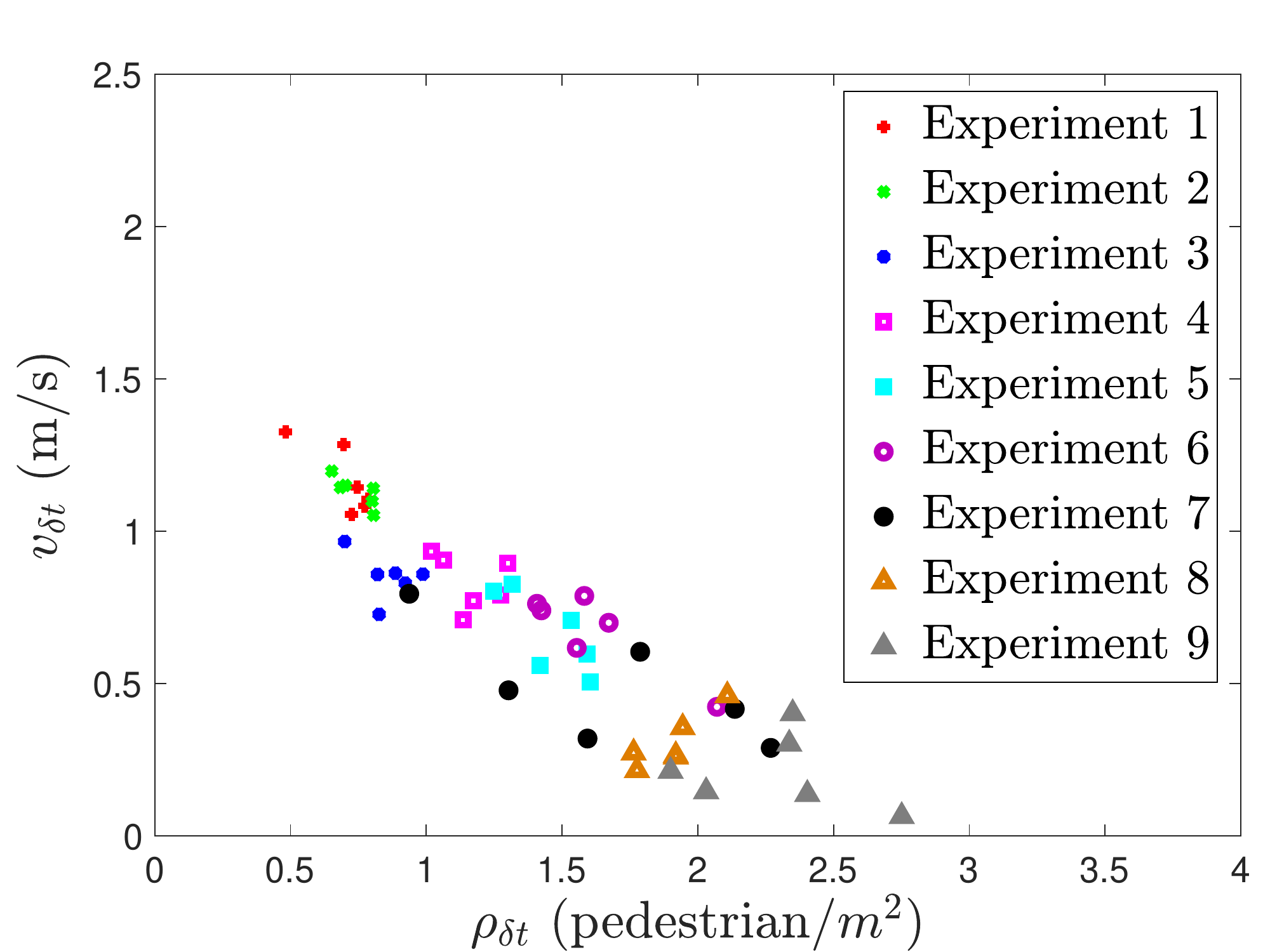}}
			\subfloat[]{\includegraphics[width=.4\textheight]{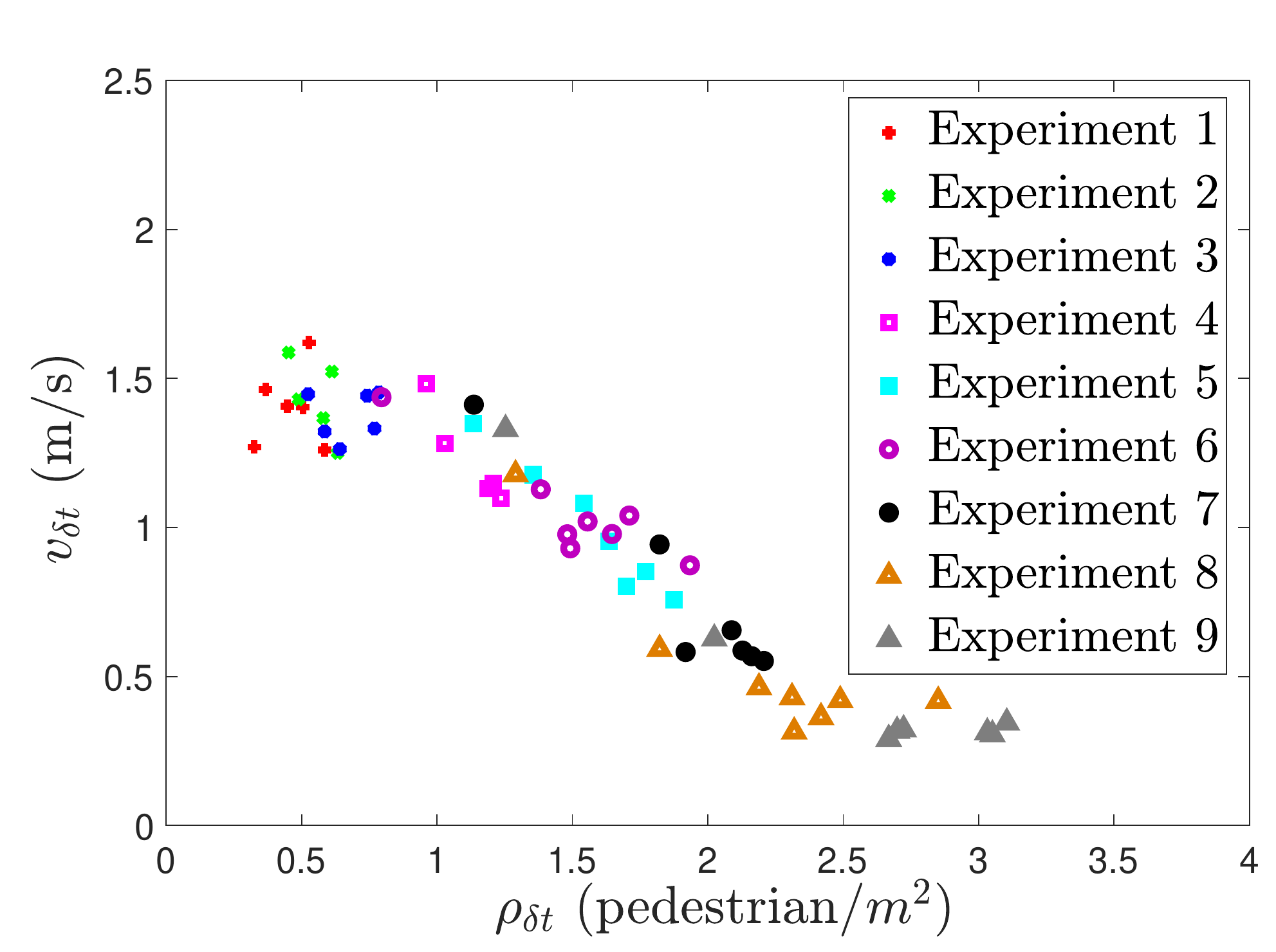}}
		\end{tabular}  
		\caption{(a) Computed and (b) measured fundamental plots for the 9 experiments in Table \ref{table:macro}. The measured data 
		are taken from \cite{Zhang2011}.}
		\label{fig:flux}
	\end{center}
\end{figure}

Fig.~\ref{fig:flux} shows the computed and measured (from \cite{Zhang2011}) fundamental diagram for each experiment in Table \ref{table:macro}. 
We observe good agreement: the overall trend is similar, although for certain experiments 
the computed average velocity for a given density is slightly smaller
than the corresponding measured quantity. We remark that the parameters in our model were set as explained 
in Sec.~\ref{subsec:Calibration} and not tuned to fit the measured data. Thus, some slight difference is to be expected. 

%%%%%%%%%%%%%%%%%%%%%%%%%%%%%%%%%%%%%%%%%%%%%%%%%%%%%%%%%%%%%%%%%%%%%%%%%%%%
\section{Bidirectional flow}
\label{sec:Laneformation}

The experimental validation of our model in Sec.~\ref{subsec:MacroValidation} involves unidirectional flow of pedestrians. 
In this section, we consider bidirectional flow. It is know that when groups of people approach each other from opposite directions, 
they form lanes (see, e.g., \cite{Helbing1995,Helbing2004180,Yu2005}), which increase the flow efficiency. 
We aim at checking that our model, calibrated as reported in Sec.~\ref{subsec:Calibration}, can reproduce this spontaneous crowd behavior. 
%
%\begin{figure}
%	\centering
%	\includegraphics[width=.6\textwidth]{lane_layout.png}
%	\caption{A schematic diagram of the corridor used as domain in lane formation experiments.}
%	\label{fig:lane_layout}
%\end{figure}

\begin{figure}[]
	\centering
	\includegraphics[height=.17\linewidth]{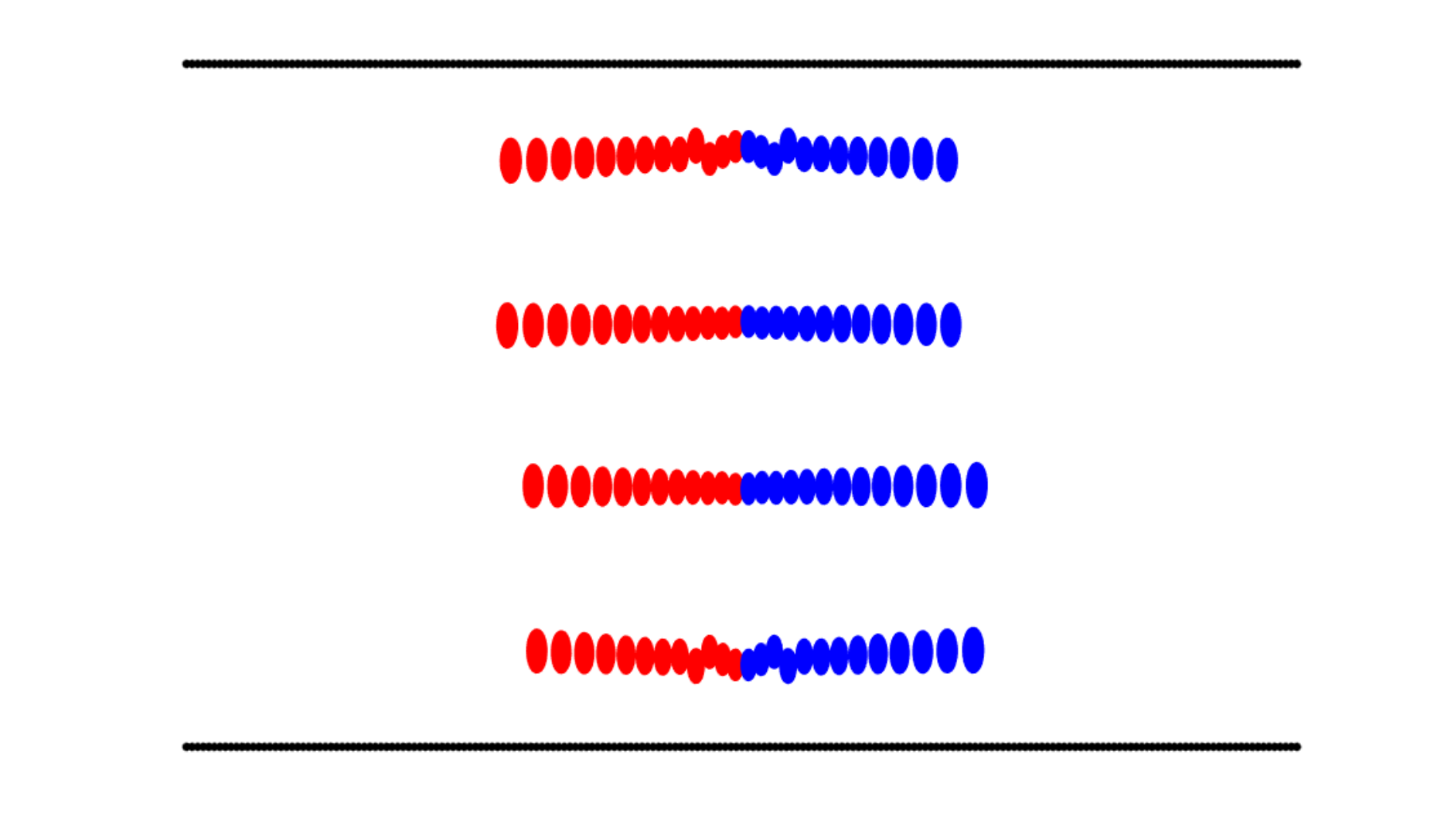}
	\caption{The ``frozen state'' of a group of 100 pedestrians initially positioned in a symmetric configuration. Red (resp., blue) pedestrians 
	move from left to right (resp., from right to left).}
	\label{fig:freezing}
\end{figure}

We consider a $20$ m long and $5$ m wide corridor. Two groups of equal size are initially placed at opposite corridor ends, 
moving towards each other. 
%We split the pedestrians into two equal groups and place them in a corridor of length (see Fig.~\ref{fig:lane_layout}). Group $1$ has the objective of moving to the right exit (target checkpoint for Group $1$ with radius 5 m) and Group $2$ has to move towards the left exit (target checkpoint for Group $2$  with radius 5 m). 
Since lane emergence is not immediate, we simulate a periodic corridor.
We vary the pedestrian density inside the corridor from $0.2 $ to $1.6$ pedestrians/m$^{2}$, with increment 
$\Delta\rho = 0.2$  pedestrians/m$^{2}$. 
Higher values of $\rho$ were not considered as they led to an unstable system in such a small geometry.
Recall that for unidirectional flow we could simulate up to $\rho = 2.75$ pedestrians/m$^{2}$ without instabilities (see Fig.~\ref{fig:flux}).
So, bidirectional flow in a confined environment seems to be challenging for our model. 
The number of pedestrians $N$ for each experiment is calculated by $N = \rho * A_c$, where 
with $A_c = 100$ m$^2$ is the corridor area. 
%We take $dt = 0.01$ s, $\tau = 0.50$ s, $d_i^0 = 0.2 $m, $t_d = 0.2$ s, $r_p = r_w = 2$ m, $\mu = 0.20$ and $\mu_w = 0.20$ for all the cases in this section. 
For all the simulation, every pedestrian has a desired speed of 1 m/s. We remark that pedestrians cannot be initially placed in a symmetric configuration in order to avoid a ``frozen state''. See Fig.~\ref{fig:freezing}. We let each simulation run until the lane configuration remains stable for $1$ minute.

\begin{figure}[]
	\begin{center}	
		\begin{tabular}{c}
			\subfloat[Two lane formation, 100 pedestrians]{\includegraphics[width=0.3\textheight]{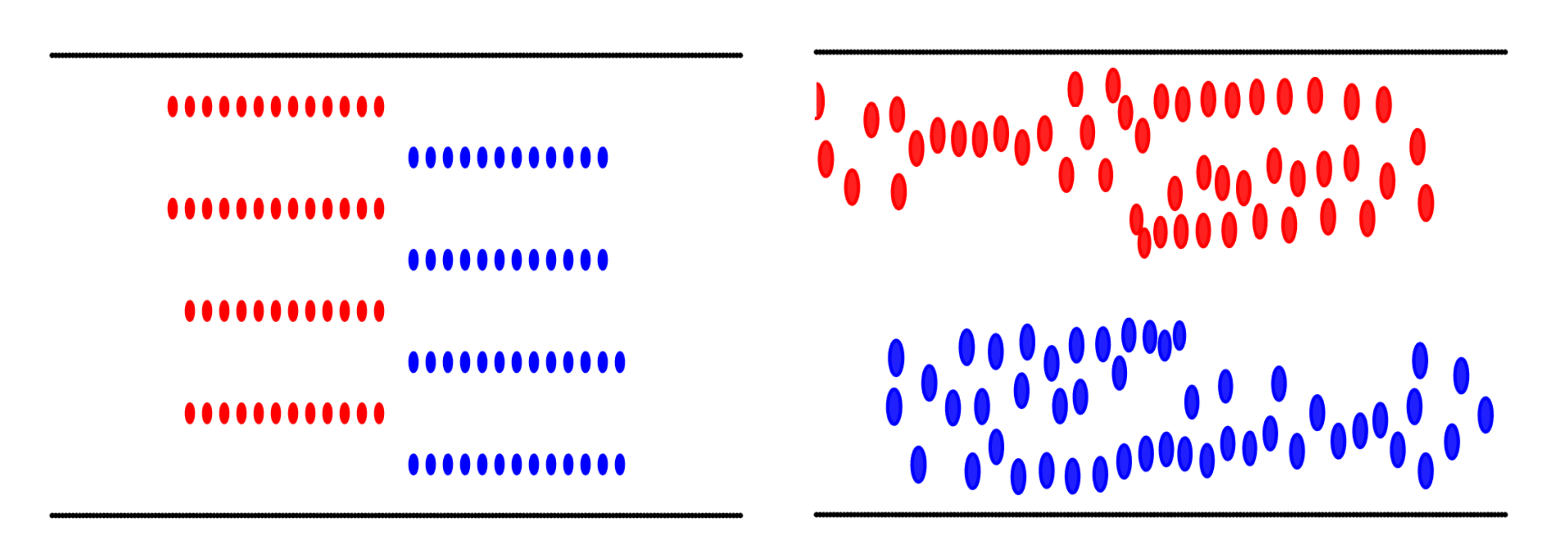}}	\quad	
			\subfloat[Four lane formation, 140 pedestrians]{\includegraphics[width=0.3\textheight]{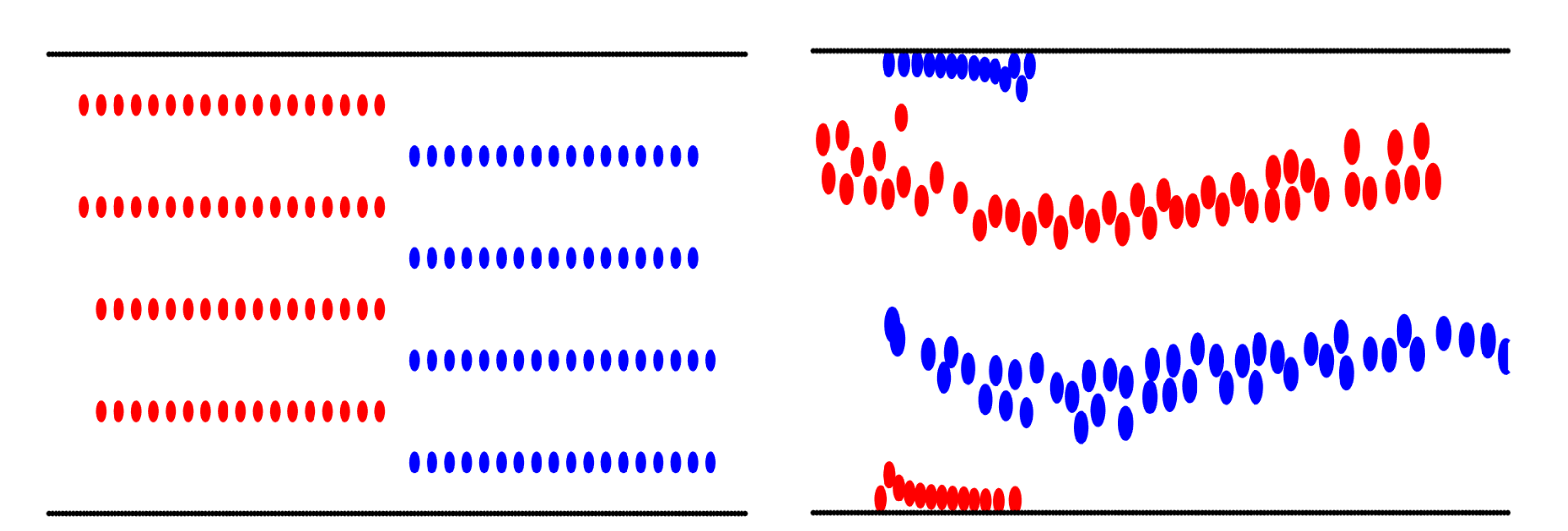}} 
		\end{tabular}  
		\caption{Initial position (left in both panels) and final lane formation configuration (right in both panel) for 
		(a) 100 people and (b) 140 people. Red (resp., blue) pedestrians 
	move from left to right (resp., from right to left).}
		\label{fig:lane_formation}
	\end{center}
\end{figure}

\begin{figure}
	\centering
	\includegraphics[width=0.3\textheight]{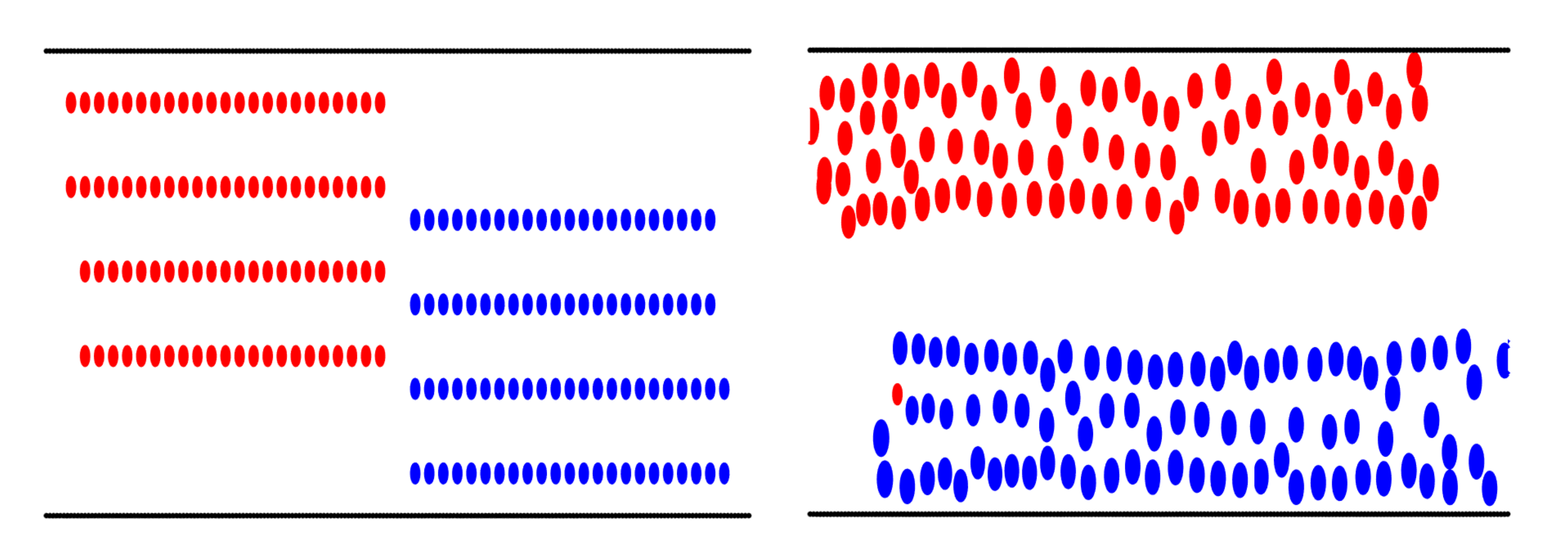}
	\caption{Initial position (left) and lane formation configuration (right) for 180 people before the system becomes unstable.
	Red (resp., blue) pedestrians move from left to right (resp., from right to left).}
	\label{fig:180_lane}
\end{figure}

Table.~\ref{table:lane_formation} reports the number of lanes formed in each experiments. We observe that 
for $\rho \leq 1.2$ two lanes form, while for $\rho = 1.4, 1.6$ four lanes form. Fig.~\ref{fig:lane_formation} 
show the lane configuration for $\rho$ = $1$ and $1.4$. In Fig.~\ref{fig:lane_formation} (b) we see that the row of pedestrians
near the wall tend to get closer to the wall when they encounter the opposing stream of pedestrians.This is due to the fact that the repulsive force from pedestrians \eqref{eq:rep_force} is
higher than the repulsive force from the walls  \eqref{eq:bou_force}.
When we considered $\rho = 1.8$, we observed two lanes before the system became unstable.
See Fig. \ref{fig:180_lane}. Notice that we used a different initial configuration than in the 
$\rho = 1.4$ case in the effort to obtain a stable system.
This suggests that the number of lanes for a short period of time is influenced 
not only by the crowd size but also by the initial positioning.

\begin{table}[]	
	\begin{center}
		\begin{tabular}{|c|c|c|c|c|c|c|c|c|}
			\hline
			Density $(\rho)$ & 0.2&0.4&0.6&0.8&1.0&1.2&1.4&1.6 \\
			\hline
			Number of pedestrians $(N)$ &20&40&60&80&100&120&140&160 \\
			\hline
			Number of lanes&2&2&2&2&2&2&4&4\\
			\hline		  
		\end{tabular}
		\caption{The number of lanes formed in each of the 8 experiments that vary by density $\rho$ and hence by number of pedestrians $N$.}	
		\label{table:lane_formation} 
	\end{center} 
\end{table} 

%%%%%%%%%%%%%%%%%%%%%%%%%%%%%%%%%%%%%%%%%%%%%%%%%%%%%%%%%%%%%%%%
% Section 6 = Contact Tracking Model definition

\section{Simulation of airborne disease spreading using pedestrian dynamics}
\label{sec:CTIntro}

With the goal of studying the onset of an airborne disease spreading in a complex, yet confined environment
we extend the model presented in Sec.~\ref{sec:Model} by introducing a simple method to trace contact.
%Contact tracing is a method for identifying all the people who could have been infected by a disease as a result of direct contact with a sick person. A sick person who could potentially infect other people with their disease/infection is termed as a $primary$  $contact$. A person is termed a $secondary$ $contact$ if they got infected by a primary contact due to sufficient contact between them. Once the secondary contacts are aware of their exposure to the disease, necessary measures could be taken to diagnose them and later treat them if needed.  
Depending on the nature of the disease, the sufficient contact time to infect a 
susceptible person varies. We will focus on a highly infectious disease, like measles. 
%The main purpose of contact tracing is to detect the early symptoms of the disease on secondary contacts, observe and treat them if possible. This process is meant to stop infections and diseases spreading further through the community and hopefully prevent an outbreak of the disease. 

\subsection{A simple model for the spreading of a highly infectious disease}
\label{subsec:CTModel}

Let us consider a group of $N$ pedestrians in a geometry $\Omega$. Pedestrians can be:  
sick (infectious), immune (non-susceptible), vulnerable (susceptible), infected and not infected.
We define the set of the susceptible and infectious people:
\begin{align}
\mathcal{S} = \{ i \in \mathbb{N}, i \leq N ~:~ i^{th} \text{ pedestrian  is  susceptible}  \}, \cl
\mathcal{I} = \{ i \in \mathbb{N}, i \leq N ~:~ i^{th} \text{ pedestrian is infectious}  \}, \el
\end{align}
respectively.
In each simulation, $P_{imm}$ $\% $ of the population is immune and only a small number of people are sick. 
The rest of the population is vulnerable. 

For a sick person $i$ with position $\br_i$, we define the set of all susceptible pedestrians that lie inside its circle of influence (sickness domain) at a certain time $t \in (0, T]$ as:
\begin{equation}
\mathcal{I}^{sus}_i = \{ j \in \mathcal{S} ~:~|| \br_j - \br_i || \leq r_s \}, \ i \in \mathcal{I},  \el
\label{eq:VulnerableSet}
\end{equation} 
where $\left \| . \right \|$ denotes the Euclidean norm in $\mathbb{R}^2$ and $r_s$ is the cutoff radius for the circle of influence. 
If a pedestrian $j$ stays in $\mathcal{I}^{sus}_i$ for a continuous period of time, e.g. $t_v$ minutes, then 
they have a $v_s$ $\% $ probability of getting infected. After $t_v$ minutes, pedestrian $j$ is moved to either one of the following sets according to its updated stage:
\begin{align}
\mathcal{E}_{sick} = \{ j \in \mathbb{N},  ~:~ j \in  \mathcal{I}^{sus}_i \ \text{for} \ t_v \ \text{mins and is infected} \},\cl
\mathcal{E}_{safe} = \{  j \in \mathbb{N},  ~:~ j \in  \mathcal{I}^{sus}_i \ \text{for} \ t_v \ \text{mins and is not infected} \}.\el
\end{align}
Once a vulnerable pedestrian moves to either of these sets, they are no longer considered to be in the
population that could get infected by sick people, i.e., 
\begin{align}
\mathcal{S} \cap \ \mathcal{E}_{sick} = \mathcal{S} \cap \ \mathcal{E}_{safe} = \emptyset. \el
\end{align} 
We also assume that pedestrians belonging to $\mathcal{E}_{sick}$ are not able to transmit the disease, although infected.
At the end of each simulation, we will have the number of secondary contacts for the simulated scenario.

In order to simulate the disease spreading in airport terminals, 
we implemented the pedestrian dynamics model in Sec.~\ref{sec:Model} and the
above contact tracing method in C++. All the simulations ran on a shared 40-core computing server with 512 GB RAM. 

%%%%%%%%%%%%%%%%%%%%%%%%%%%%%%%%%%%%%%%%%%%%%%%%%%%%%%%%%%%%%%%%
% Section 7 = Numerical Results of Contact Traking  

\subsection{Numerical Results}
\label{sec:CTNumericalResults}

We consider terminals in two US airports, Hobby Airport in Houston and the Atlanta International Airport, as concrete settings
to test our pedestrian dynamics model with contact tracing. We will examine
different scenarios: variable population size, variable percentage of immune (non-susceptible) population, 
boarding bridges or boarding buses. 
Through the numerical simulations, we quantify the increase in average number of secondary contacts increases
as a given terminal becomes more densely populated, the percentage of immune people decreases,
the number of primary contacts increases, and areas of high density (such as the boarding buses) are present.

Since we deal with more complicated geometries that the ones tested in Sec.~\ref{sec:Model} and \ref{sec:Laneformation}, 
we handle the checkpoint assignment in a different way. The list of checkpoints contains, for a given checkpoint $i$, 
the checkpoint position and a radius $r_i$. If checkpoint $i$ is in a person's path, the position of the checkpoint 
for that particular person is randomly picked to be a point inside the circle of radius $r_i$ centered at checkpoint $i$'s position.
This makes each pedestrian's checkpoints unique (if truly random). 
In addition, this way to assign checkpoints helps to avoid oscillations in pedestrian motion due to clustering of  
pedestrians around checkpoints that have wait times.

%%%%%%%%%%%%%%%%%%%%%%%%%%%%%%%%%%%%%%%%%%%%%%%%%%%%%%%%%%%%%%%%%%%%5

\subsection{Hobby Airport in Houston}
\label{subsec:Houston}

\begin{figure}[]
	\centering
	\includegraphics[width=.7\textwidth]{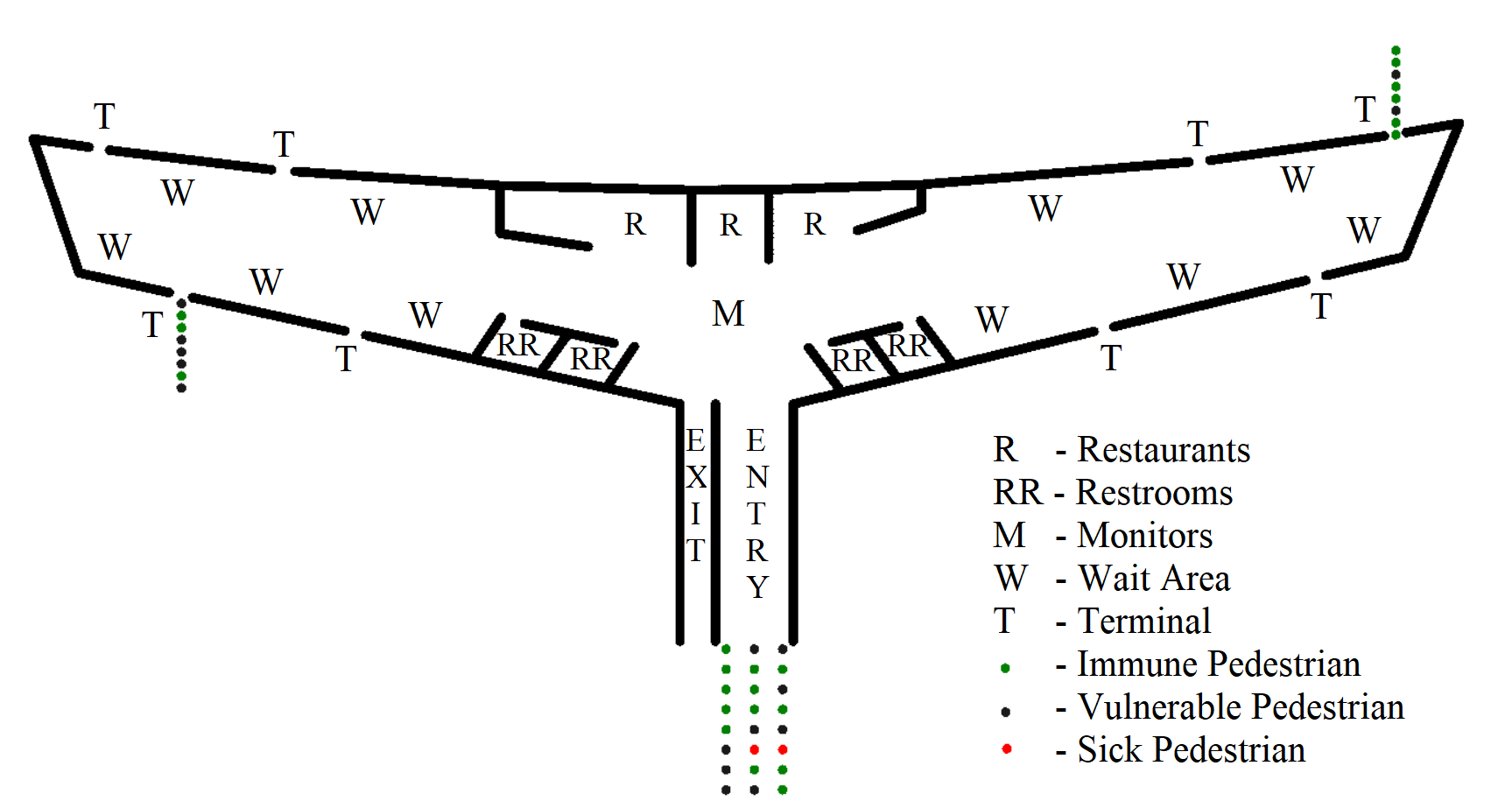}
	\caption{Part of the William P. Hobby Airport in Houston (USA). Pedestrians are represented with a dot.}
	\label{fig:airport}
\end{figure}

We consider a part of Houston's William P. Hobby (HOU) Airport as a sample geometry $\Omega$. See Fig.~\ref{fig:airport}.
At the start, each of the $N$ pedestrian in $\Omega$ is randomly categorized to be either sick, immune or vulnerable. 
Pedestrians are assigned a random path to pass through the airport. Some people deplane, enter the 
airport through the terminal gate and leave the airport via the \emph{exit} corridor. Others enter the airport through the \emph{entry} 
corridor and walk to their assigned terminal gate. Random people are selected to use the restrooms or stop at a restaurant. 
Departing people are also assigned to randomly check display monitors. 
Appropriate wait times are allocated for each checkpoint that denotes a restroom, restaurant or a display monitor.
Finally, if a person reaches the gate before their boarding time, they stay in the \emph{wait area} near their assigned gate 
until it is time to board. 

The pedestrian dynamics model parameters are set as follows: $dt = 0.01$ s, $\tau = 0.5$ s, $\tau_d = 0.18$ s, $r_p = r_w = 2$ m, $d_i^0 = 0.20$ m, and $\mu = \mu_w = 0.3$. For the contact tracing, we set $r_{s} = 2.5$ m 
and $v_s = 90 \%$. The latter is a realistic value for a highly infectious disease. $P_{imm}$ will vary from $90 \%$ to $55 \%$ with $\Delta P_{imm} =  -5 \%$. Remark that herd immunity for a highly infectious disease, like measles, is over 90\%. For each $P_{imm}$, 200 simulations are run to calculate the average number of secondary contacts, which is denoted by $Avg_{sc}$. 

We will consider a simple case to test our implementation (case 1) and a more realistic case (case 2):
\begin{itemize}
\item[-] \textbf{Case 1:} $N$ = 400, $T$ = $15$ minutes, $t_v$ = 1 minute.
\item[-] \textbf{Case 2:} $N$ = 1000, $T$ = $60$ minutes,  $t_v$ = $2$ minutes.
\end{itemize}
For each case, we consider two scenarios: a) 1 primary contact and b) 2 primary contacts. We remark that both cases feature a low pedestrian 
density, given the size of the domain. As expected, our model can handle them efficiently.

\begin{figure}[h]
	\centering
	\includegraphics[width=0.7\textwidth]{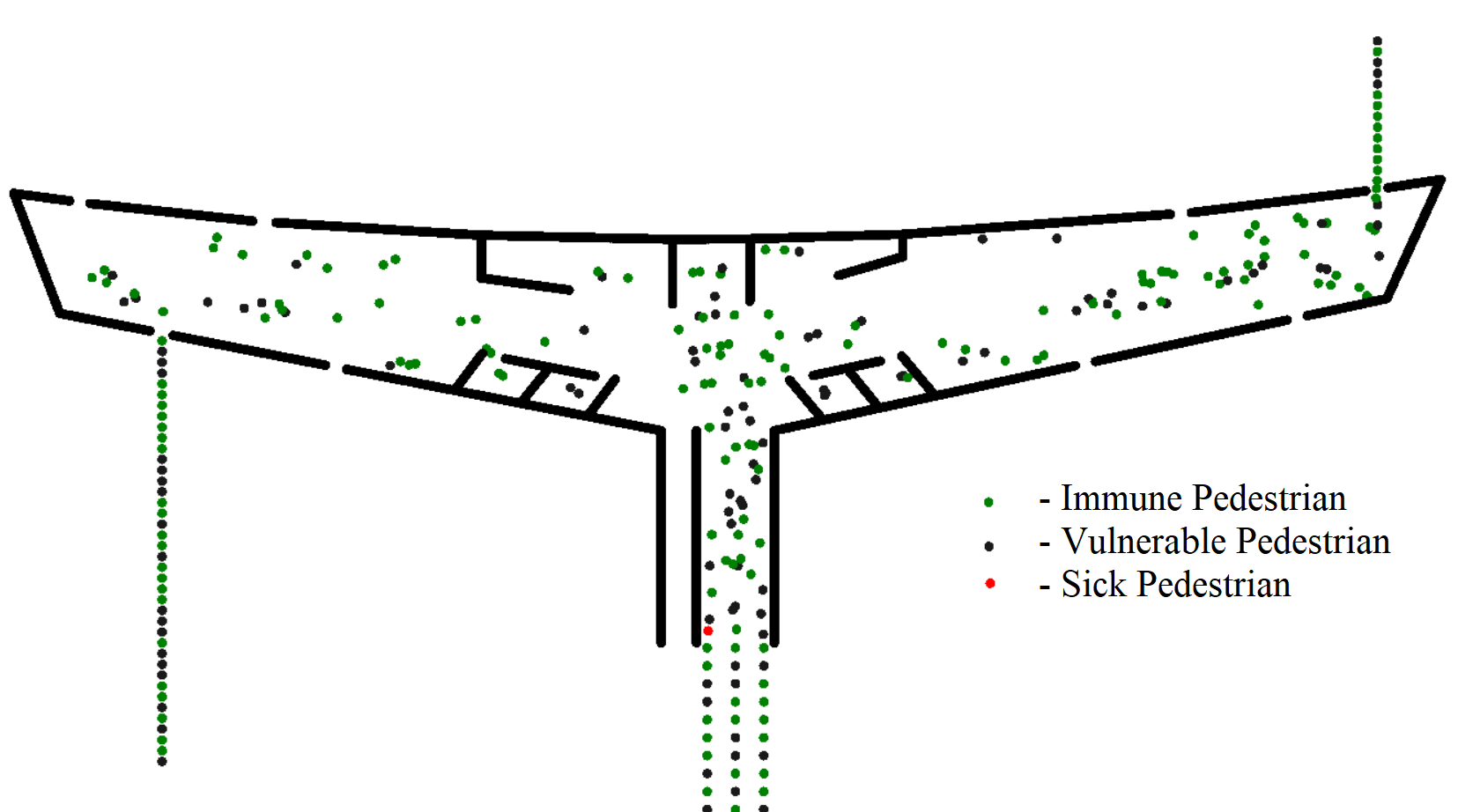}
	\includegraphics[width=.7\textwidth]{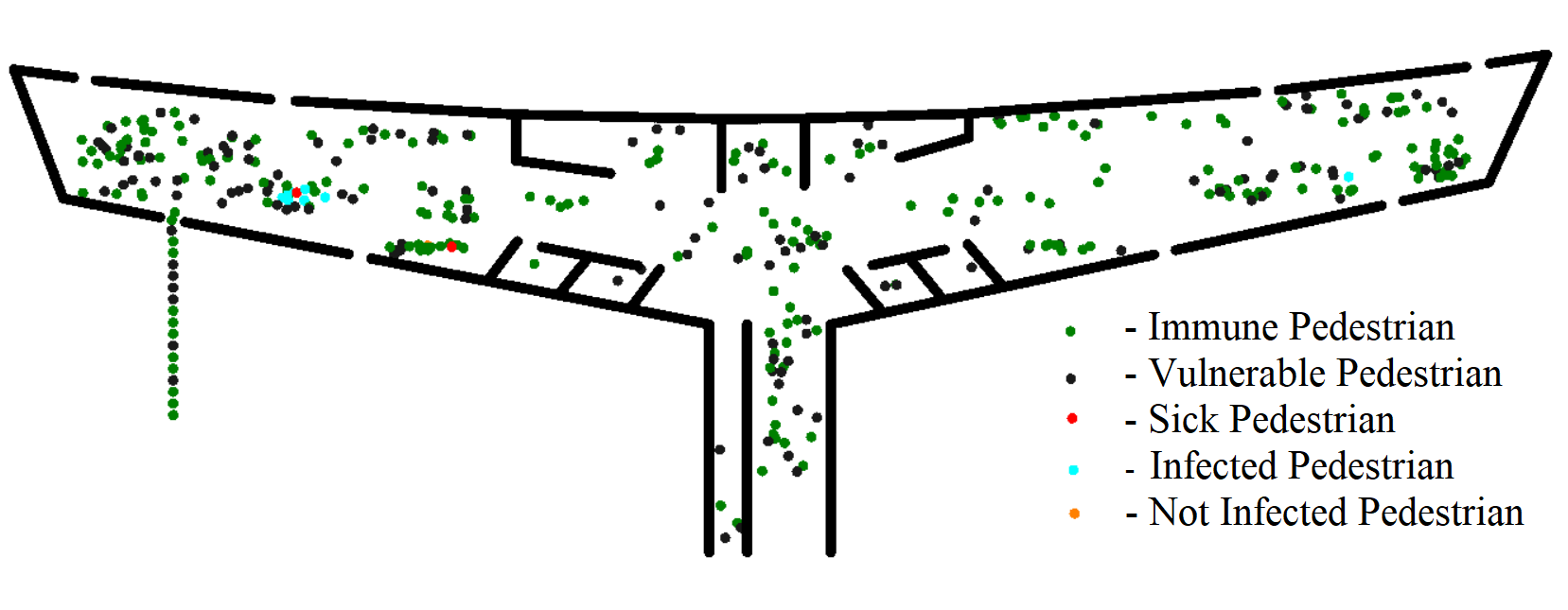}
	\caption{HOU, case 1b: people distribution and characterization after 5 minutes from the start (top) and at the end of the simulation (bottom), 
	i.e.~after 12 minutes.}
	\label{fig:airport_initial_final}
\end{figure}

Fig.~\ref{fig:airport_initial_final} shows two snapshots of a simulation for case 1b. 
Dots denote people and the color refers to their characterization: red for sick (primary contact), 
green for immune, black for vulnerable, cyan for infected (secondary contact), and orange for 
not infected. In Fig.~\ref{fig:airport_initial_final} (top), people are deplaning from the rightmost gate while people
at the leftmost gate are waiting to deplane. 
In Fig.~\ref{fig:airport_initial_final} (bottom), we see that the primary contact infected several surrounding people
in the waiting area. 

\begin{figure}[]
	\centering
	\includegraphics[width=0.5\textheight]{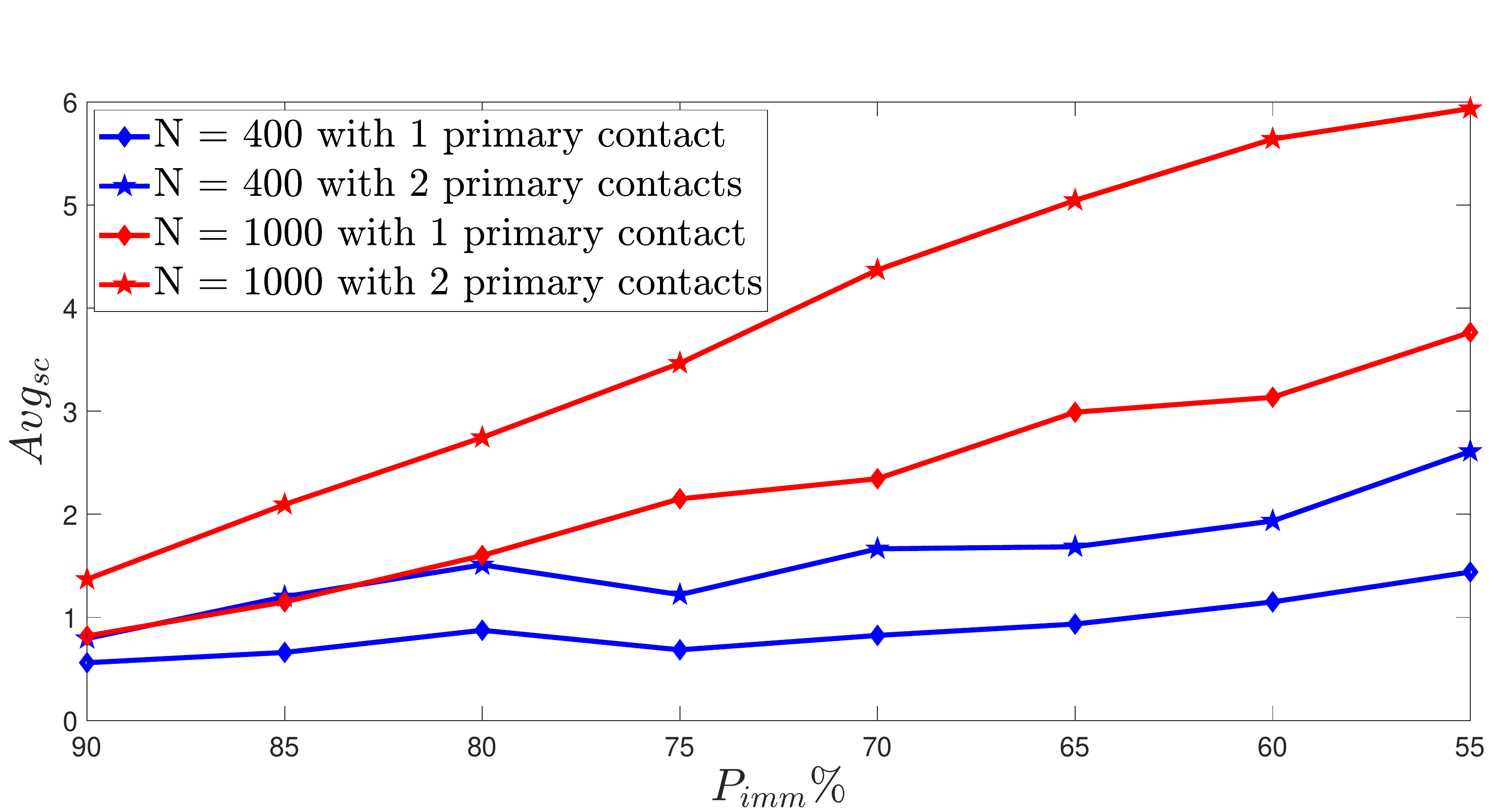}
	\caption{HOU: average number of secondary contacts $Avg_{sc}$ as the population size $N$ and the percentage
	of immune pedestrians $P_{imm}$ varies.}
	\label{fig:no_buses}
\end{figure}

Fig.~\ref{fig:no_buses} shows the average number of secondary contacts for varying $P_{imm}$. We note that 
with the exception of $P_{imm}$ = $75 \%$ for case 1, when $P_{imm}$ decreases the number of secondary contacts
increases for both cases. We also note that $Avg_{sc}$ for case 1 increases slowly as $P_{imm}$ decreases, while 
the rate of increase of $Avg_{sc}$ is faster for case 2. 

\begin{figure}
	\begin{center}	
	\begin{tabular}{ c }
		\subfloat[Case 1a: $N$ = 400, 1 primary contact.]{\includegraphics[width=0.49\textwidth]{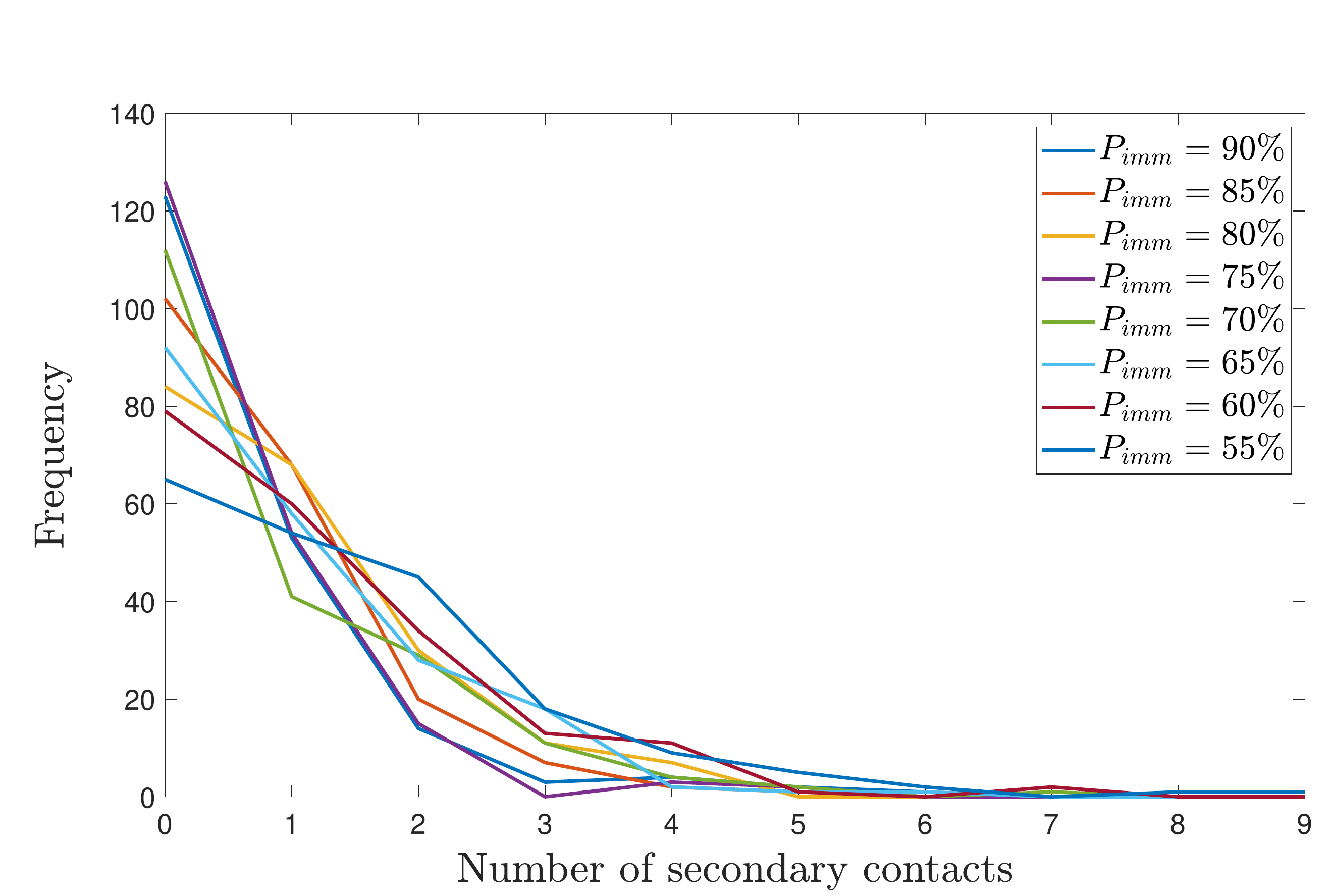}}	
		\subfloat[Case 2b: $N$ = 1000, 2 primary contacts.]{\includegraphics[width=.49\textwidth]{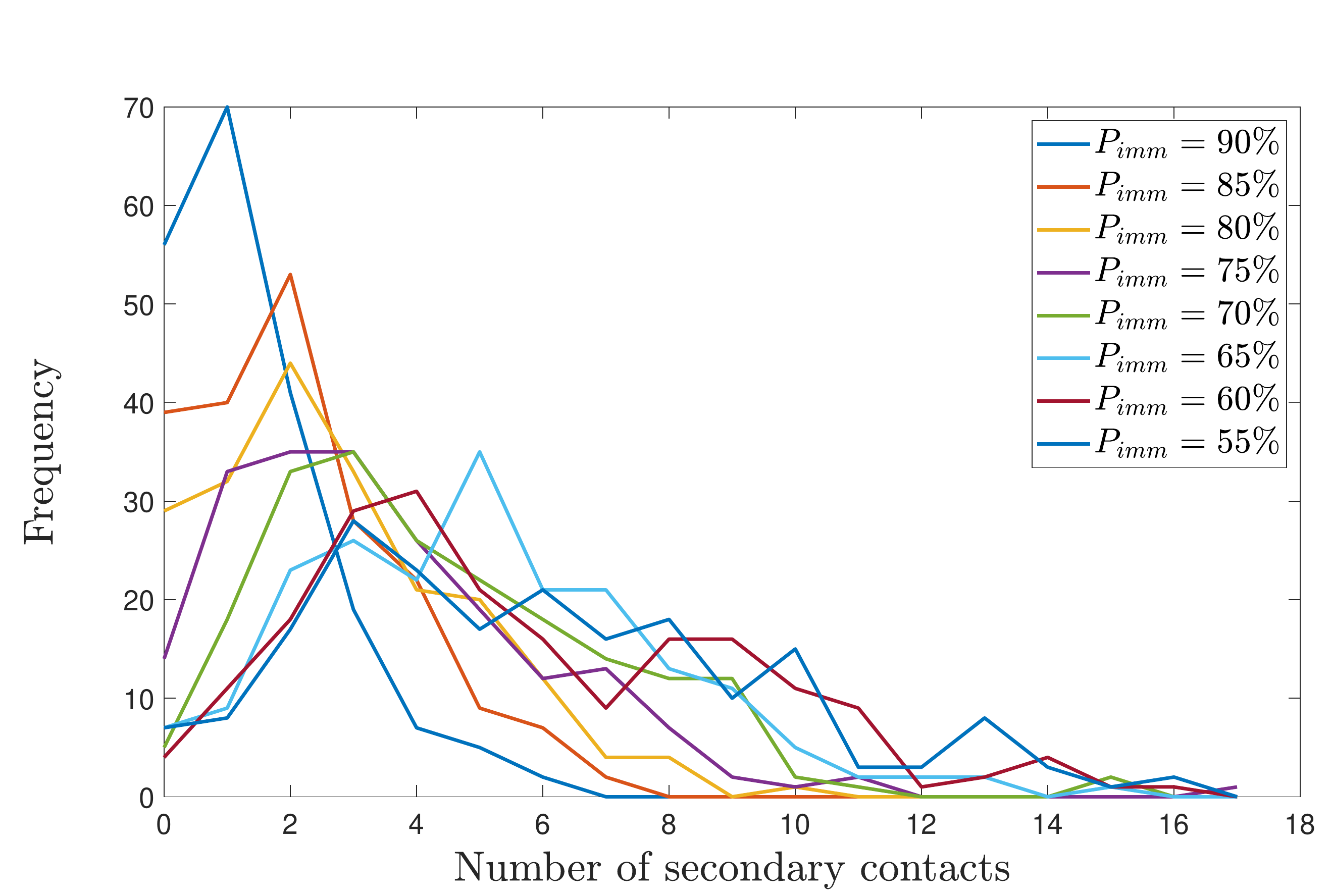}} 
	\end{tabular}   
	\caption{HOU: frequency distribution of the number of secondary contacts produced in 200 simulations
	for (a) case 1a and (b) case 2b for varying $P_{imm}$.}
	\label{fig:freq}
\end{center}
\end{figure}

Table~\ref{table:400_1sick} reports the computed average number of secondary contacts 
$Avg_{sc}$, standard deviation $Sd_{sc}$ for all 200 simulations and maximum number of secondary contacts 
$Max_{sc}$ among all the simulations for cases 1a and 2b, respectively. We see that as $P_{imm}$ decreases, 
the standard deviation of the number of secondary contacts increases, indicating a wider range of 
values for the number of secondary contacts. This is reinforced by Fig.~\ref{fig:freq}, where 
we compare the frequency distributions for cases 1a and 2b.
The curves for larger $P_{imm}$ values have taller, narrower peaks as opposed to the curves for smaller $P_{imm}$ values.
From Fig.~\ref{fig:freq} (a) we see that the mode of every distribution is zero, while the mode of every distribution in Fig.~\ref{fig:freq} (b) is different from zero. 

\begin{table}
	\begin{center}
		\begin{tabular}{|c|c|c|c|c|c|c|c|c|}
		\hline
		\multicolumn{9}{|c|}{Case 1a} \\
			\hline
			$P_{imm} \%$ &90&85&80&75&70&65&60&55 \\
			\hline
			$Avg_{sc}$ &0.56&0.66&0.88&0.69&0.83&0.94&1.15&1.44 \\
			\hline
			$Sd_{sc}$ &0.93&0.85&0.99&1.19&1.19&1.12&1.31&1.53 \\
			\hline
			$Max_{sc}$ &4&4&4&7&7&6&7&9 \\
			\hline	
			\multicolumn{9}{|c|}{Case 2b} \\
			\hline	  
			$P_{immune} \%$ &90&85&80&75&70&65&60&55 \\
			\hline
			$Avg_{sc}$ &1.37&2.10&2.75&3.47&4.37&5.05&5.64&5.94 \\
			\hline
			$Sd_{sc}$ &1.30&1.66&2.04&2.53&2.69&2.80&3.37&3.57 \\
			\hline
			$Max_{sc}$ &6&7&10&17&15&15&16&16 \\
			\hline	
		\end{tabular}
		\caption{HOU, case 1a and 2b: computed average number of secondary contacts $Avg_{sc}$, 
		standard deviation $Sd_{sc}$ for all 200 
		simulations and maximum number of secondary contacts $Max_{sc}$ among all the simulations.}	
		\label{table:400_1sick} 
	\end{center} 
\end{table}

%%%%%%%%%%%%%%%%%%%%%%%%%%%%%%%%%%%%%%%%%%%%%%%%%%%%%%%%%%%%%%%%%%%%%%%%%%%%

%\subsubsection{Airport terminal with buses to board the plane}
%\label{subsubsec:Houstonwithbuses}

\begin{figure}[h]
	\centering
	\includegraphics[width=0.7\textwidth]{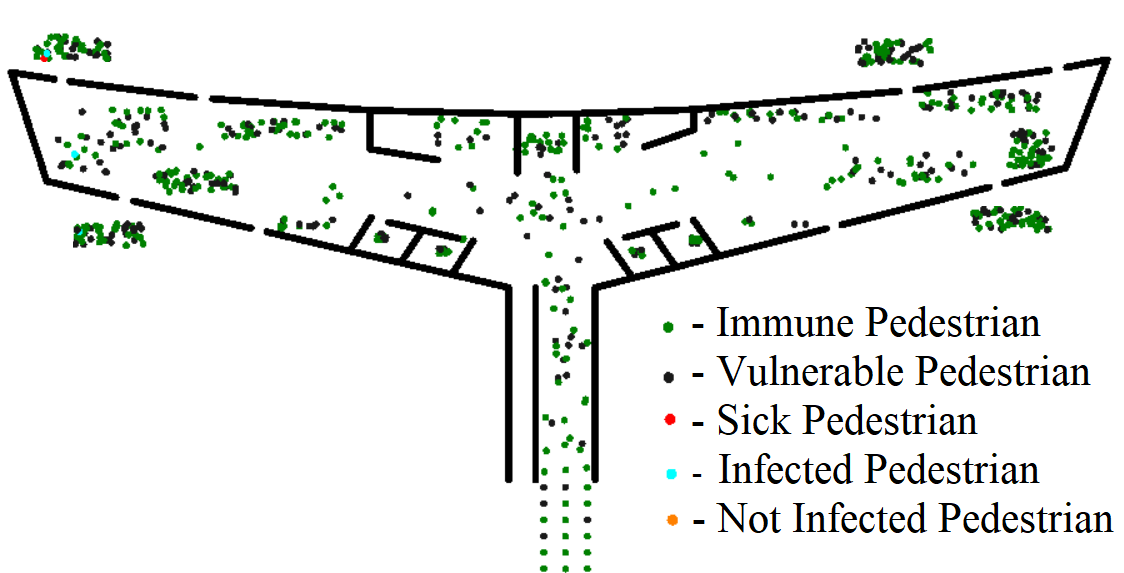}
	\caption{HOU with buses: screenshot of a simulation. Pedestrians are represented with a dot and color encodes
	the different characterization.}
	\label{fig:hobby_buses}
\end{figure}

Although at HOU airport passengers board through a boarding bridge, we extend the geometry in Fig.~\ref{fig:airport}
to include buses to transport passengers from the gate to the plane. This is to compare the disease spreading in 
an airport with and without buses. 
Every departing pedestrian will have their final checkpoint as a random position inside a bus. The pedestrians stay in the bus
for 5 minutes and then they board the plane. A bus is 5 m x 12 m and can accommodate 50 people.
Fig.~\ref{fig:hobby_buses} shows a screenshot of a simulation in HOU Airport with buses. 

\begin{figure}[h]
	\begin{center}	
		\begin{tabular}{c c }
			\subfloat[without buses]{\includegraphics[width=0.48\textwidth]{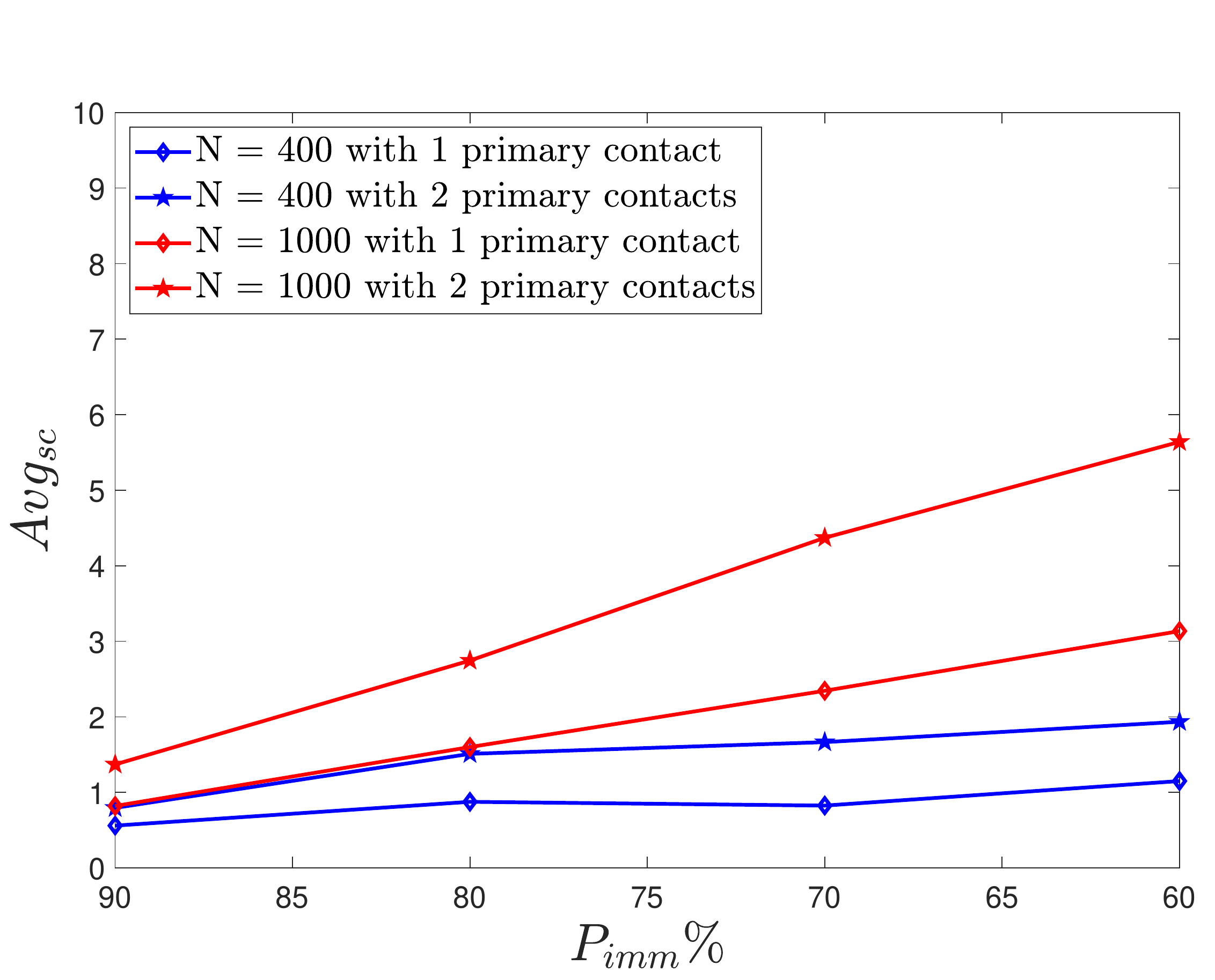}}	&	
			\subfloat[with buses]{\includegraphics[width=.48\textwidth]{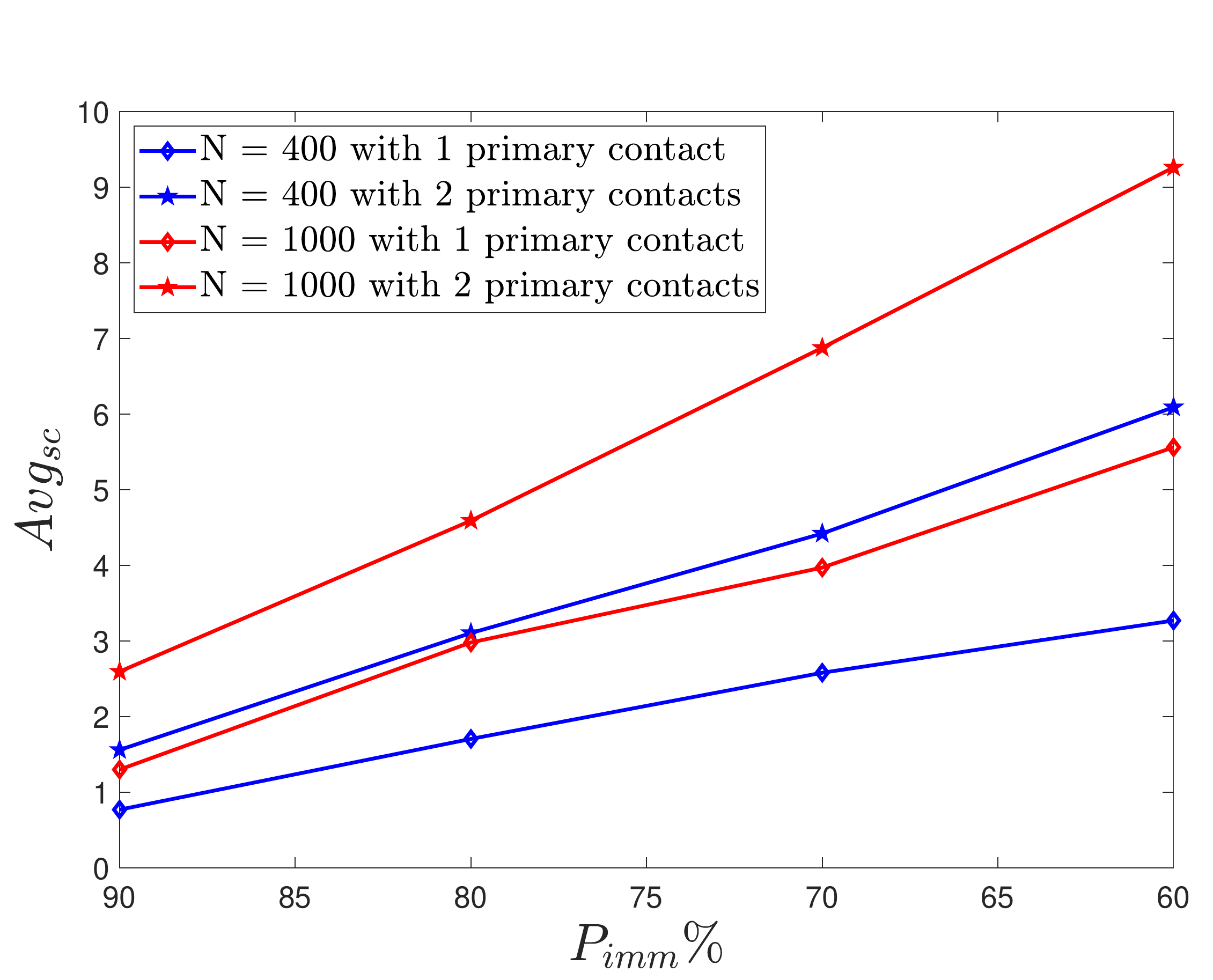}} 
		\end{tabular}   
		\caption{HOU: average number of secondary contacts $Avg_{sc}$ produced (a) without buses
		and (b) with buses as the percentage of immune pedestrians $P_{imm}$ varies.}
		\label{fig:avg_combined}
	\end{center}
\end{figure}

\begin{figure}[h]
	\centering
	\includegraphics[width=.6\textwidth]{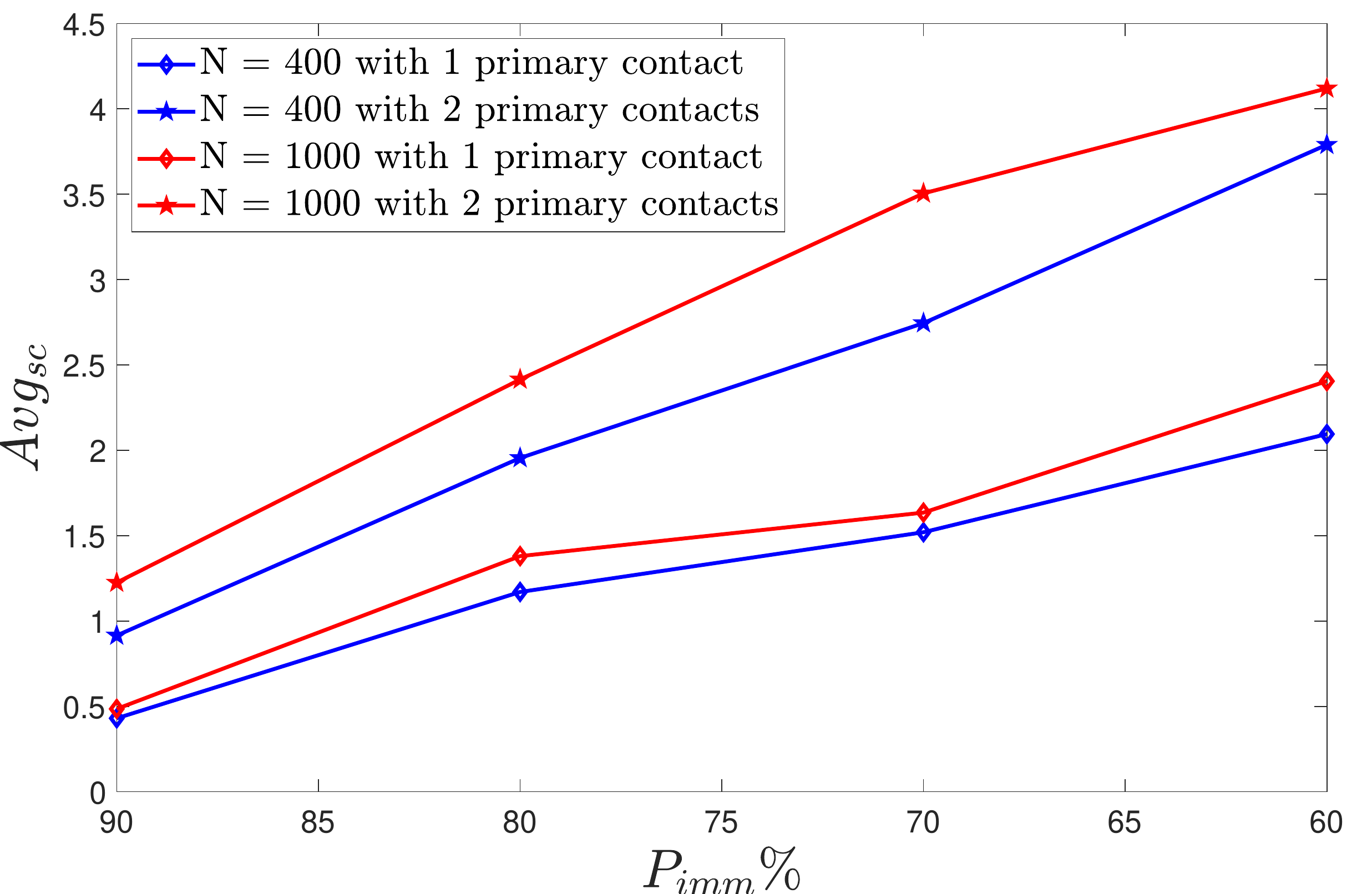}
	\caption{HOU: average number of secondary contacts generated inside the buses  
	as the percentage of immune pedestrians $P_{imm}$ varies.}
	\label{fig:only_buses}
\end{figure}

We run 200 simulations, with the parameters set as in the case of no buses, in order to calculate $Avg_{sc}$ for $P_{imm}$ = $90 \%$, 
$80 \%$, $70 \%$, $60 \%$. Fig.~\ref{fig:avg_combined} shows $Avg_{sc}$ produced in HOU Airport with and without buses. 
The data in Fig.~\ref{fig:avg_combined} (a) is the same as in Fig.~\ref{fig:no_buses}, but with a different scale to facilitate the comparison with Fig.~\ref{fig:avg_combined} (b). We see that having a high density area such as a bus in the path 
of pedestrians drastically increases the rate at which people get infected, for a given value of $P_{imm}$. In addition, 
the rate at which $Avg_{sc}$ increases as $P_{imm}$ decreases is faster in the airport with buses. 
We report in Fig.~\ref{fig:only_buses} the average number of secondary contacts generated inside the bus. 
We observe that the cases with the same number of primary contacts 
have average values closer to each other. The population size $N$ makes less of a difference
because the number of people in the buses does not change significantly between case 1 and 2. 

Finally, to estimate the basic reproduction number, we set $P_{imm} = 0\%$ for cases 1a and 2a.
Table \ref{Table:R0_Houston} reports the basic reproduction number averaged over 200 simulations
for both cases without and with buses. For both cases, the buses contribute to an 
increase of roughly 5 infected people.
  
\begin{table}
 	\begin{center}
 		\begin{tabular}{|c|c|c|}
 			\hline
 			& Without Buses & With Buses \\
 			\hline
 			case 1a & 2.4 & 7.66 \\ 
 			\hline
 			case 2a & 9.305 & 15.21 \\ 
 			\hline
 		\end{tabular}
 	\caption{HOU: average basic reproduction number without and with buses.}
 	\label{Table:R0_Houston}
 	\end{center}	
 \end{table}
 
%%%%%%%%%%%%%%%%%%%%%%%%%%%%%%%%%%%%%%%%%%%%%%%%%%%%%%%

\subsection{Connecting airports: Hobby Airport to Atlanta International Airport}

\begin{figure}
	\centering
	\includegraphics[width=0.5\textwidth]{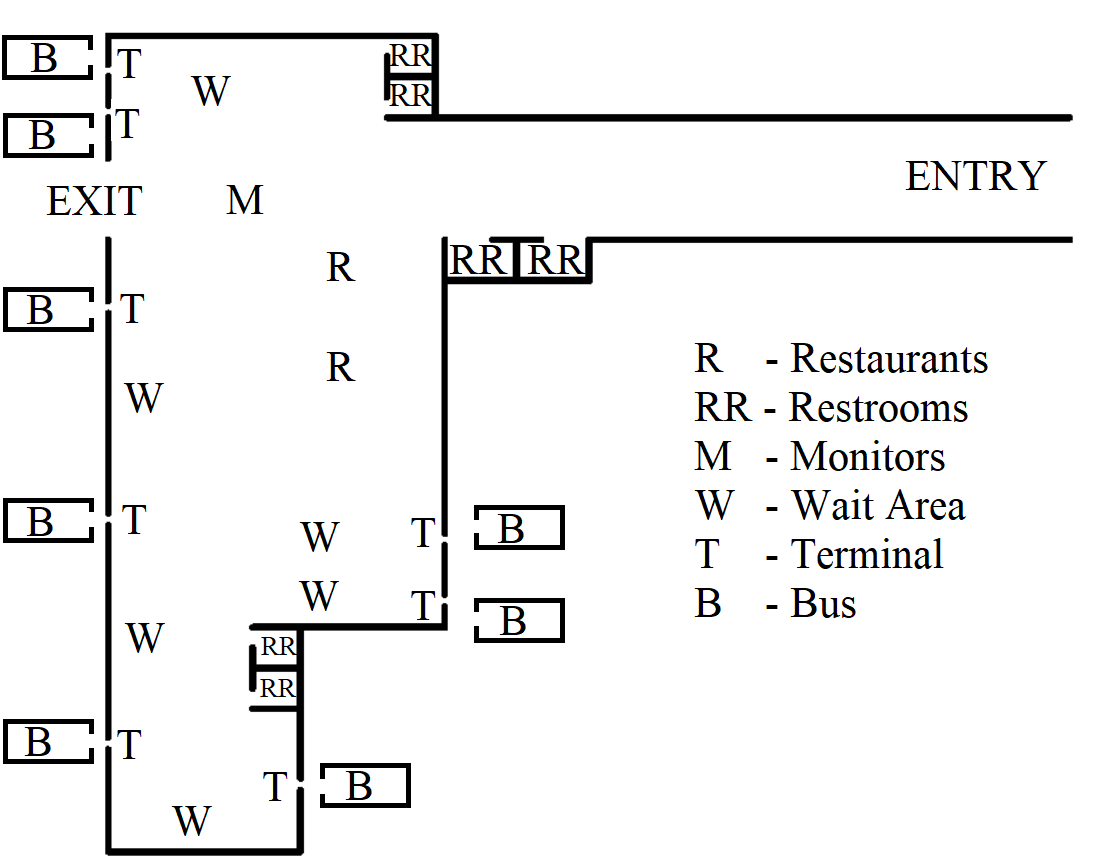}
	\caption{Part of the Hartsfield-Jackson Atlanta International Airport in Atlanta (USA) used for the simulations.}
	\label{fig:Atlanta}
\end{figure}

The set of tests presented in this section involves one or two primary contacts
that takes a flight from Hobby Airport (shown in Fig.~\ref{fig:airport})
to Hartsfield-Jackson  Atlanta International (ATL) Airport. We will only consider the part of ATL airport shown
in Fig.~\ref{fig:Atlanta}. In all the simulations, the primary contact(s) 
enters HOU airport through security check, moves through the terminal to reach the gate of their
flight, boards the plane to ATL airport. Once at ATL airport, the primary contact(s)
moves through the terminal to reach the gate of their final flight. 
Just like for the results in Sec.~\ref{subsec:Houston}, some people enter ATL airport through the entry corridor and board a flight,
while others enter through the gates. Ten percent of the latter group goes onto boarding a flight, while the rest 
leave the airport through the exit. In addition, random people are selected to use the restrooms or stop at a restaurant and
departing people are also assigned to randomly check display monitors.
We will consider the case of both airports with and without buses.

All the parameters are set like in Sec.~\ref{subsec:Houston}. However, we limit the values of $P_{imm}$ to 
$90 \%$, $80 \%$, $70 \%$, $60 \%$ and we consider only one case: $N$ = 1000, $T$ = $50$ minutes,  $t_v$ = $2$ minutes, with
a) 1 primary contact and b) 2 primary contacts.
For each scenario, we run 200 simulations and calculate the average number of secondary contacts $Avg_{sc}$.

\begin{figure}
	\begin{center}	
		\begin{tabular}{c c }
			\subfloat[$t = 0$]{\includegraphics[width=0.45\textwidth]{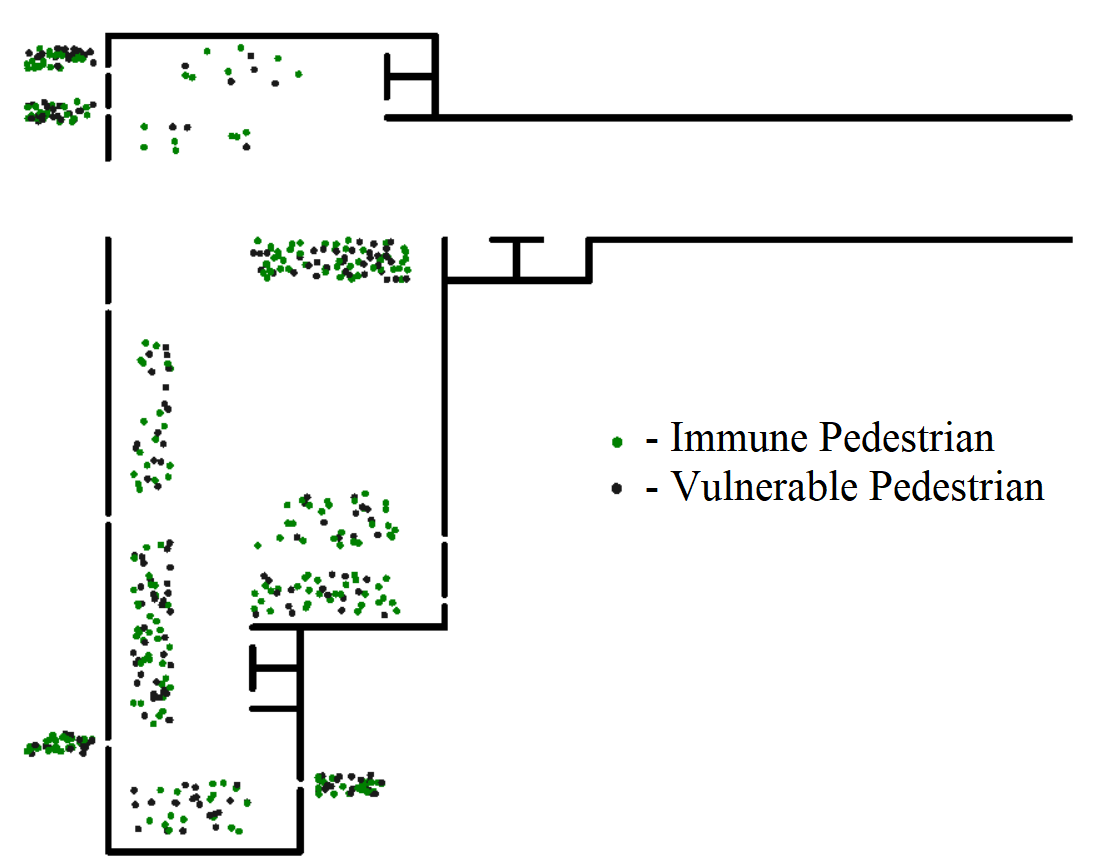}}	&	
			\subfloat[$t = 20$ minutes]{\includegraphics[width=0.48\textwidth]{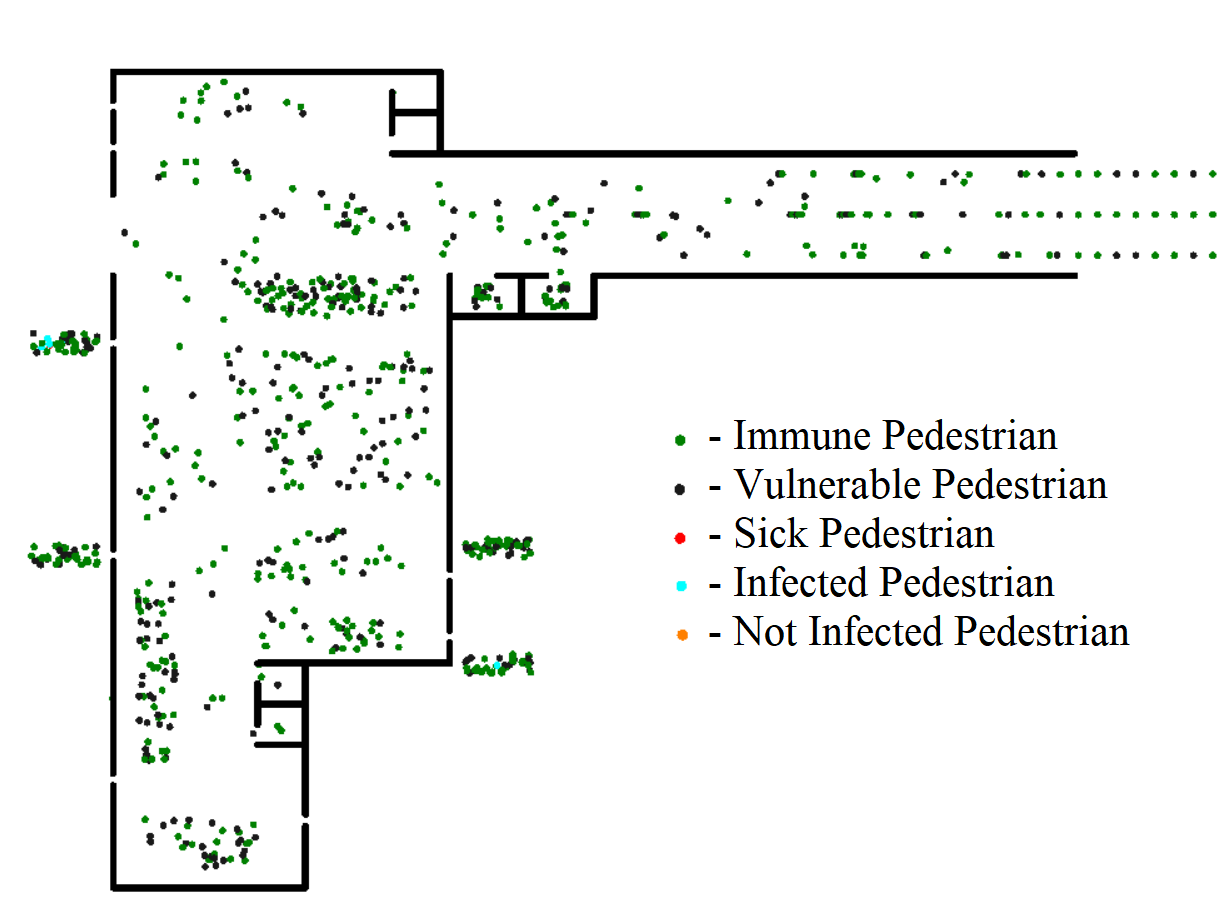}} 
		\end{tabular}   
		\caption{
		ATL, case 1a: (a) initial people distribution and characterization and (b) after 20 minutes.}
		\label{fig:Atlanta_initial_final}
	\end{center}
\end{figure}

To simplify the simulation, the sick person can infect vulnerable people only inside the terminals and, if present, in the buses. 
In this way, the simulations in each airport can be run in parallel.
Fig.~\ref{fig:Atlanta_initial_final} (a) shows the initial people distribution for one of the simulations 
in ATL airport with the buses. 
The pedestrians inside the buses spend 5 minutes there before they reach the terminal. The people in the terminal
are waiting to board their planes. The rest of the pedestrians, including the primary contact(s), 
will enter the airport later as per their assigned arrival time 
(if deplaning) or $15$ minutes into the simulation (if entering through the entry corridor). 
%\kri{We set that the primary contact always entered ATL airport by deplaning, they then spend time inside the airport and then they board a plane again. We did this as we were planning to initially have one more airport!}
Fig.~\ref{fig:Atlanta_initial_final} (b) shows the people distribution and characterization after 20 minutes.

\begin{figure}[h]
	\begin{center}	
		\begin{tabular}{ c c }
			\subfloat[Without buses.]{\includegraphics[width=0.48\textwidth]{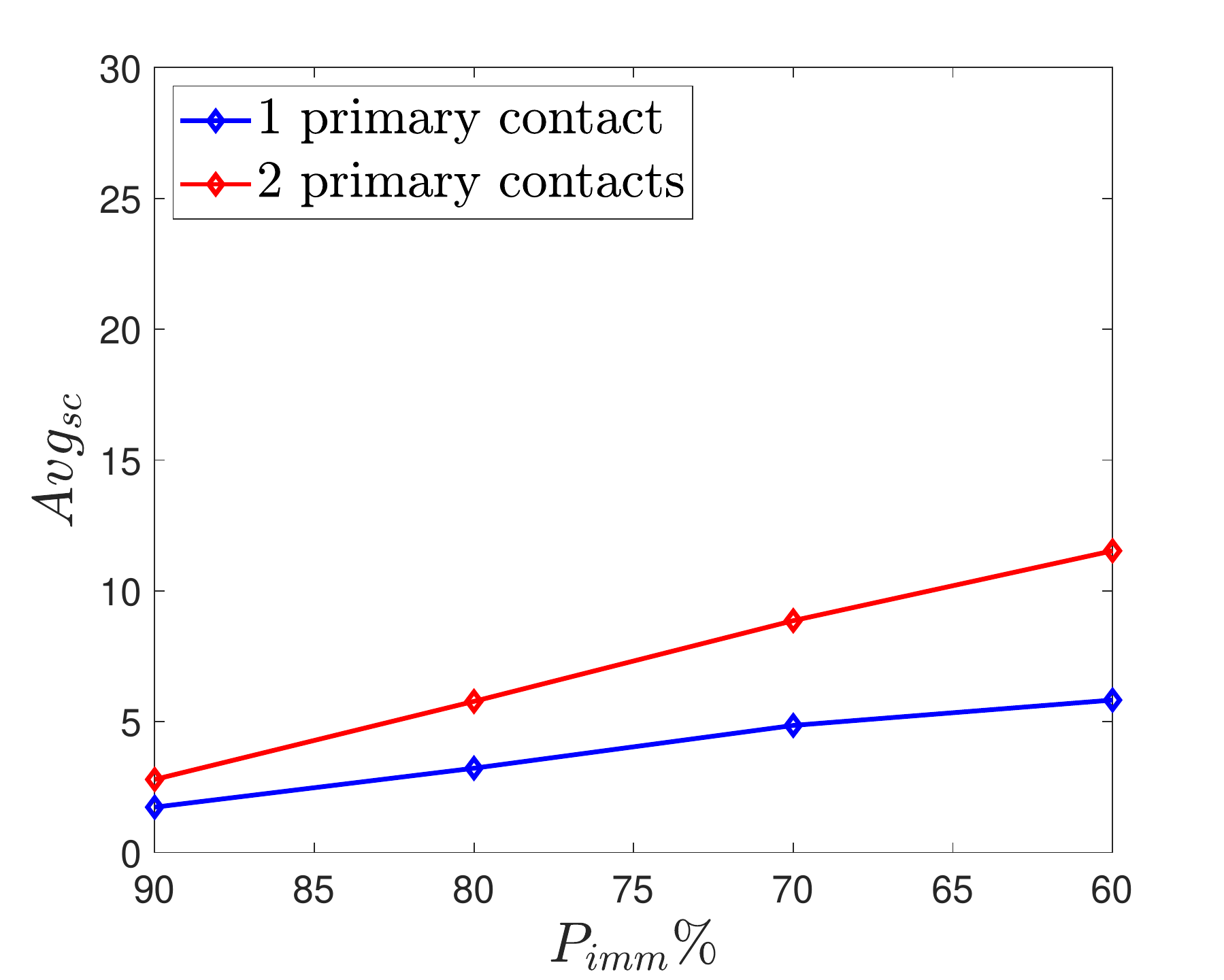}}	&	
			\subfloat[With buses.]{\includegraphics[width=0.48\textwidth]{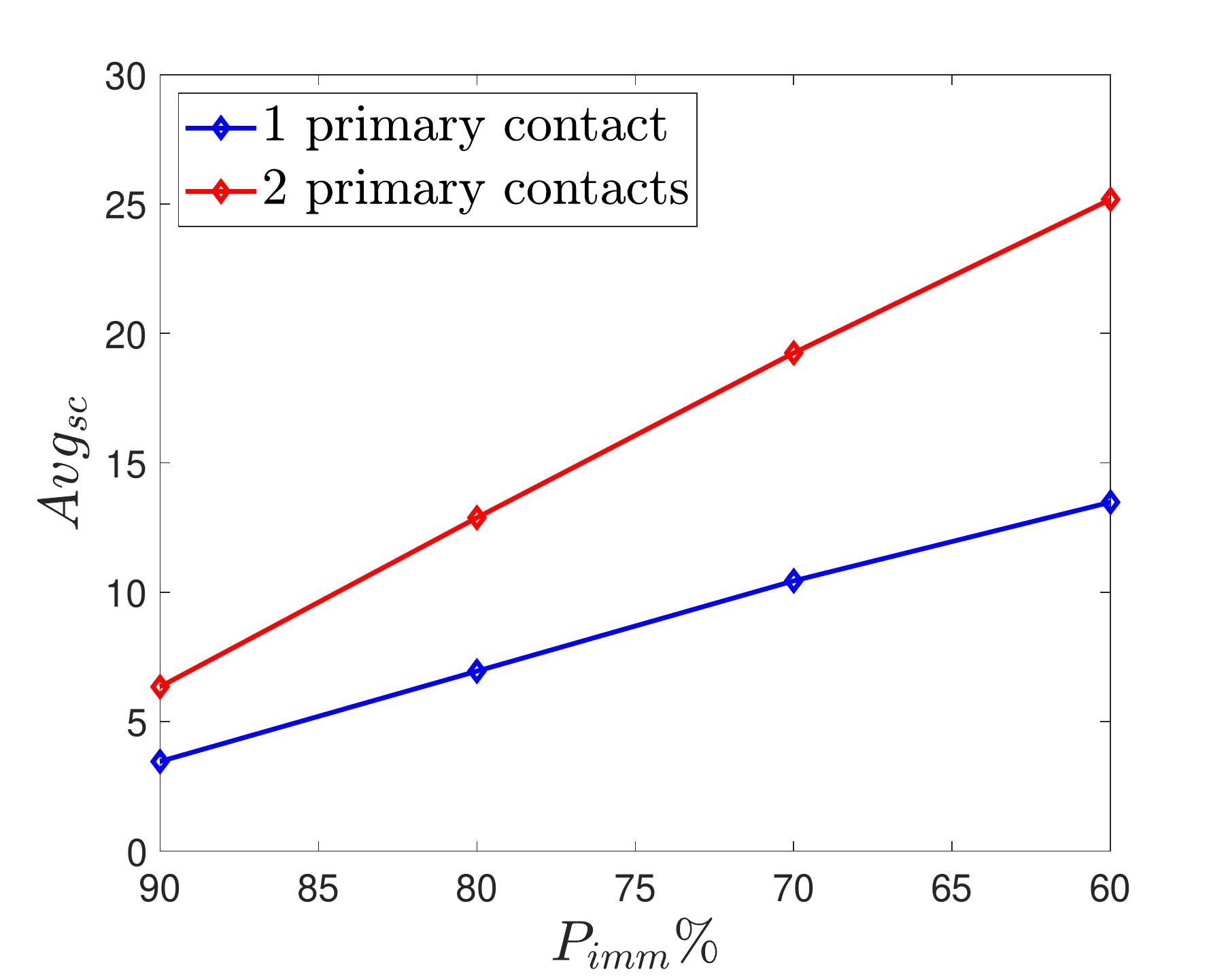}} 
		\end{tabular}   
		\caption{HOU to ATL (a) with buses or (b) without buses: average number of secondary contacts produced by 
		1 or 2 primary contacts for different values of $P_{imm}$. Each airport contains 1000  people.}
		\label{fig:Final_avg}
	\end{center}
\end{figure}

Fig.~\ref{fig:Final_avg} reports the the average number of secondary contacts for case 1a and 1b with and without buses. 
By comparing Fig.~\ref{fig:Final_avg} (a) and (b), we see that the number of secondary contacts is more than doubled
in presence of the buses. 
These results confirm that: (i) a high percentage of immune pedestrians in the system ensures that the 
number of infected people remain small and (ii) the spreading of the disease is amplified by the airport buses.  

\begin{figure}
	\centering
	\includegraphics[width=0.48\textwidth]{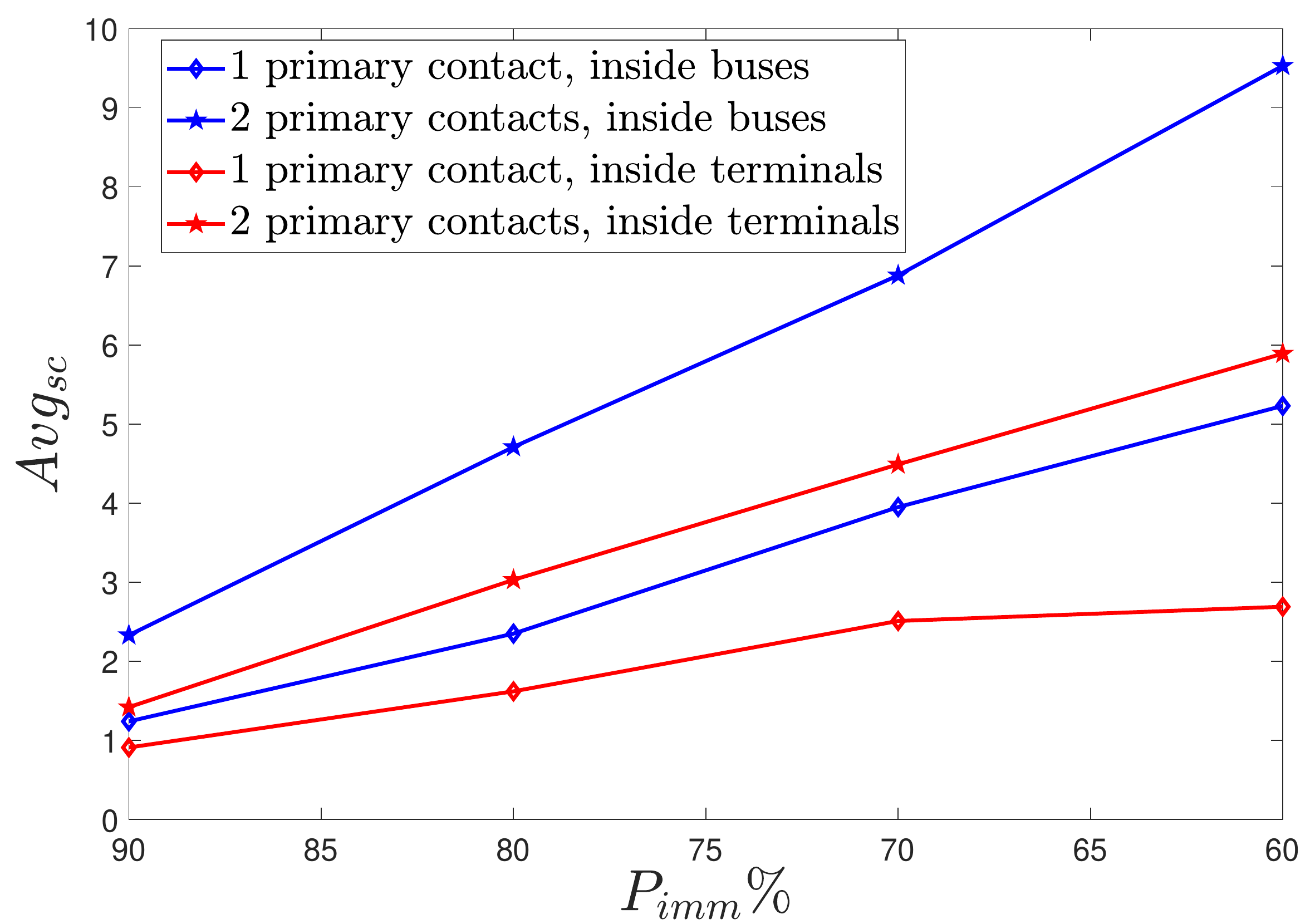}
	\caption{ATL, cases 1a and 1b: average number of secondary contacts produced for different $P_{imm}$ 
	in specific locations (only inside buses vs only inside the terminal).}
	\label{fig:Atl_derived}
\end{figure}

Fig.~\ref{fig:Atl_derived} displays the average number of secondary contacts produced inside the buses vs inside the 
ATL terminal for cases 1a and 1b. We notice that the average number of people infected by one primary contact inside the buses 
is close to the average number of people infected by two primary contacts inside the terminal.

Finally, to estimate the basic reproduction number in ATL airport we set $P_{imm} = 0\%$ for case 1a.
Table \ref{Table:R0_Atlanta} reports the basic reproduction number averaged 
over 200 simulations for cases 1a without and with buses.	
		\begin{table}[]
		\begin{center}
			\begin{tabular}{|c|c|c|}
				\hline
				& Without Buses & With Buses \\
				\hline
				Case 1a & 7.263 & 18.693 \\ 
				\hline
			\end{tabular}
			\caption{Average basic reproduction number in the geometry in Fig.~\ref{fig:Atlanta} without and with buses.}
		\label{Table:R0_Atlanta}
		\end{center}	
	\end{table}

%%%%%%%%%%%%%%%%%%%%%%%%%%%%%%%%%%%%%%%%%%%%%%%%%%%%%%%%%%%%%%%%
% Section 8 = Conclusion

\section{Conclusions}
\label{sec:Conclusion}

We combined a grid free, force-based microscopic model for pedestrian dynamics with a contact tracing
method to study the initial spreading of a highly infectious airborne disease in a confined environment. 
The pedestrian dynamics model is calibrated to avoid unrealistic pedestrian motion (overlap of people
and oscillating walking trajectories) and validated against empirical data from \cite{Zhang2011}. 
In addition, we showed that lane formation in bidirectional flow (a well known self organizing phenomenon)
is captured by the pedestrian dynamics model under consideration. 
The contact tracing method uses a sickness domain around each sick person. 
A susceptible person has a prescribed probability to become infected if they stay in a sickness domain for a certain amount of time. 

To test our model of pedestrian dynamics with contact tracing, we considered medium size populations with immune and susceptible 
people in the terminals of two US airports. 
We computed the average number of secondary contacts produced by one or two primary contacts in a single terminal
or in terminals of different airports connected by a flight hosting a primary contact. 
We also considered the case where people could board a bus to reach their planes. 
Sick people were traced only when inside airports or buses. 
In the case of airports without buses, we concluded  that:
\begin{itemize}
	\item[-] For same size populations, an increase in primary contacts causes an increase in the average number of infected pedestrians
	that becomes larger as the percentage of immune population decreases.
	\item[-] The larger the population size, the higher the rate of increase in the average number of infected pedestrians. 
	\item[-] A higher population density leads to an increase in average number of secondary contacts for a fix percentage of immune population.
\end{itemize}   
In the case of airports with buses (high density areas), we showed the drastic increase in the rate at which people are getting infected. 

The combination of pedestrian dynamics model with contact tracing method could be used in different
settings (emergency rooms, hospitals, etc.) and tailored to airborne diseases with different spreading 
mechanisms. If further tested and validated, it could become a tool to investigate
best practices that can help reduce the spreading of a disease.

\section*{Acknowledgements}
This work has been partially supported by NSF through grant DMS-1620384. We acknowledge 
fruitful discussions with Drs. Ilya Timofeyev and William Fitzgibbon. 

%%%%%%%%%%%%%%%%%%%%%%%%%%%%%%%%%%%%%%%%%%%%%%%%%%%%%%%%%%%%%%%%

\bibliography{ContactTracing}

\begin{thebibliography}{10}

\bibitem{Agnelli2015}
J.~P. Agnelli, F.~Colasuonno, and D.~Knopoff.
\newblock A kinetic theory approach to the dynamics of crowd evacuation from
  bounded domains.
\newblock {\em Mathematical Models and Methods in Applied Sciences},
  25(01):109--129, 2015.

\bibitem{Ball2011}
Frank~G. Ball, Edward~S. Knock, and Philip~D. O'Neill.
\newblock Threshold behaviour of emerging epidemics featuring contact tracing.
\newblock {\em Advances in Applied Probability}, 43(4):1048--1065, 2011.

\bibitem{Bellomo2013}
N.~Bellomo, D.~Knopoff, and J.~Soler.
\newblock On the difficult interplay between life, ``complexity'';, and
  mathematical sciences.
\newblock {\em Mathematical Models and Methods in Applied Sciences},
  23(10):1861--1913, 2013.

\bibitem{Bellomo2012}
N.~Bellomo, B.~Piccoli, and A.~Tosin.
\newblock Modeling crowd dynamics from a complex system viewpoint.
\newblock {\em Mathematical Models and Methods in Applied Sciences},
  22(supp02):1230004, 2012.

\bibitem{Bellomo2011383}
Nicola Bellomo and Abdelghani Bellouquid.
\newblock On the modeling of crowd dynamics: Looking at the beautiful shapes of
  swarms.
\newblock {\em Networks and Heterogeneous Media}, 6(3):383--399, 2011.

\bibitem{Bellomo2017_book}
Nicola Bellomo, Abdelghani Bellouquid, Livio Gibelli, and Nisrine Outada.
\newblock {\em A quest towards a mathematical theory of living systems}.
\newblock Modeling and Simulation in Science, Engineering and Technology.
  Birkhauser, 2017.

\bibitem{Bellomo2013_new}
Nicola Bellomo, Abdelghani Bellouquid, and Damian Knopoff.
\newblock From the microscale to collective crowd dynamics.
\newblock {\em SIAM Multiscale Model. Simul.}, 11:943--963, 2013.

\bibitem{Bellomo2011}
Nicola Bellomo and Christian Dogbe.
\newblock On the modeling of traffic and crowds: A survey of models,
  speculations, and perspectives.
\newblock {\em SIAM Review}, 53(3):409--463, 2011.

\bibitem{Bellomo2015_new}
Nicola Bellomo and Livio Gibelli.
\newblock Toward a mathematical theory of behavioral-social dynamics for
  pedestrian crowds.
\newblock {\em Math. Models Methods Appl. Sci.}, 25(13):2417--2437, 2015.

\bibitem{Bellomo2016_new}
Nicola Bellomo and Livio Gibelli.
\newblock Behavioral crowds: {Modeling and Monte Carlo} simulations toward
  validation.
\newblock {\em Computers and Fluids}, 141:13--21, 2016.

\bibitem{Bellomo2019_new}
Nicola Bellomo, Livio Gibelli, and Nisrine Outada.
\newblock On the interplay between behavioral dynamics and social interactions
  in human crowds.
\newblock {\em Kinetic and Related Models}, 12(2):397--409, 2019.

\bibitem{Browne2015}
Glenn~Webb Cameron~Browne, Hayriye~Gulbudak.
\newblock Modeling contact tracing in outbreaks with application to ebola.
\newblock {\em Journal of Theoretical Biology}, 384:33--49, 2015.

\bibitem{Chraibi2009}
M.~Chraibi, A.~Seyfried, A.~Schadschneider, and W.~Mackens.
\newblock Quantitative description of pedestrian dynamics with a force-based
  model.
\newblock In {\em IEEE/WIC/ACM International Joint Conference on Web
  Intelligence and Intelligent Agent Technology IEEE Computer Society, Los
  Alamitos, CA}, volume~3, pages 583--586, 2009.

\bibitem{Chraibi2011425}
Mohcine Chraibi, Ulrich Kemloh, Andreas Schadschneider, and Armin Seyfried.
\newblock Force-based models of pedestrian dynamics.
\newblock {\em Networks and Heterogeneous Media}, 6(3):425--442, 2011.

\bibitem{Chraibi2010}
Mohcine Chraibi, Armin Seyfried, and Andreas Schadschneider.
\newblock Generalized centrifugal-force model for pedestrian dynamics.
\newblock {\em Phys. Rev. E}, 82:046111, Oct 2010.

\bibitem{Eichner2003}
Martin Eichner.
\newblock Case isolation and contact tracing can prevent the spread of
  smallpox.
\newblock {\em American Journal of Epidemiology}, 158:118–128, 2003.

\bibitem{Fraser2004}
C~Fraser, S~Riley, R.M Anderson, and Ferguson N.M.
\newblock Factors that make an infectious disease outbreak controllable.
\newblock {\em Proc. Natl Acad. Sci. USA}, 101:6146–6151, 2004.

\bibitem{Garnett1996}
Anderson~R.M Garnett~G.P.
\newblock Sexually transmitted diseases and sexual behavior: insights from
  mathematical models.
\newblock {\em J. Infect. Dis.}, 174:S150–S161, 1996.

\bibitem{Guzzetta2015}
Giorgio Guzzetta, Marco Ajelli, Zhenhua Yang, Leonard~N. Mukasa, Naveen Patil,
  DeniseE. Kirschner, and Stefano Merler.
\newblock Effectiveness of contact investigations for tuberculosis control in
  {Arkansas}.
\newblock {\em Journal of Theoretical Biology}, 380:238–246, 2015.

\bibitem{BS:BS3830360405}
Dirk Helbing.
\newblock A mathematical model for the behavior of pedestrians.
\newblock {\em Behavioral Science}, 36(4):298--310, 1991.

\bibitem{Helbing2004180}
Dirk Helbing.
\newblock Collective phenomena and states in traffic and self-driven
  many-particle systems.
\newblock {\em Computational Materials Science}, 30(1�2):180 -- 187, 2004.
\newblock Selected papers of the Twelfth International Workshop on
  Computational Materials Science (CMS2002).

\bibitem{Helbing1995}
Dirk Helbing and P\'eter Moln\'ar.
\newblock Social force model for pedestrian dynamics.
\newblock {\em Phys. Rev. E}, 51:4282--4286, May 1995.

\bibitem{1367-2630-1-1-313}
Dirk Helbing and Tamas Vicsek.
\newblock Optimal self-organization.
\newblock {\em New Journal of Physics}, 1(1):13, 1999.

\bibitem{Hethcote1984}
Yorke~J.A Hethcote~H.W.
\newblock Gonorrhea transmission dynamics and control.
\newblock {\em Springer Lecture Notes in Biomathematics, Berlin:Springer},
  1984.

\bibitem{HUGHES2002507}
Roger~L. Hughes.
\newblock A continuum theory for the flow of pedestrians.
\newblock {\em Transportation Research Part B: Methodological}, 36(6):507 --
  535, 2002.

\bibitem{Hyman2003}
James~M. Hyman, Jia Li, and E.~Ann, Stanley.
\newblock Modeling the impact of random screening and contact tracing in
  reducing the spread of {HIV}.
\newblock {\em Mathematical Biosciences}, 181:17--54, 2003.

\bibitem{JOHANSSON_HELBING}
Anders Johansson, Dirk Helbing, and Pradyumn~K. Shukla.
\newblock Specification of the social force pedestrian model by evolutionary
  adjustment to video tracking data.
\newblock {\em Advances in Complex Systems}, 10(supp02):271--288, 2007.

\bibitem{Kermack1927}
McKendrick~A.G Kermack~W.O.
\newblock A contribution to the mathematical theory of epidemics.
\newblock {\em Proc. R. Soc.}, 115:700–721, 1927.

\bibitem{KimQuaini2019}
Daewa Kim and Annalisa Quaini.
\newblock A kinetic theory approach to model pedestrian dynamics in bounded
  domains with obstacles.
\newblock {\em Kinetic and Related Models}, 12(6):1273--1296, 2019.

\bibitem{Kiss2005}
Istvan~Z Kiss, Darren~M Green, and Rowland~R Kao.
\newblock Disease contact tracing in random and clustered networks.
\newblock {\em Proceedings of the Royal Society B}, 272:1407–1414, 2005.

\bibitem{Klinkenberg2006}
Don Klinkenberg, Christophe Fraser, and Hans Heesterbeek.
\newblock The effectiveness of contact tracing in emerging epidemics.
\newblock {\em PLOS ONE}, 1(1):1--7, 12 2006.

\bibitem{Lakoba01052005}
Taras~I. Lakoba, D.~J. Kaup, and Neal~M. Finkelstein.
\newblock Modifications of the {Helbing-Molnar-Farkas-Vicsek} social force
  model for pedestrian evolution.
\newblock {\em SIMULATION}, 81(5):339--352, 2005.

\bibitem{6701214}
S.~Liu, S.~Lo, J.~Ma, and W.~Wang.
\newblock An agent-based microscopic pedestrian flow simulation model for
  pedestrian traffic problems.
\newblock {\em IEEE Transactions on Intelligent Transportation Systems},
  15(3):992--1001, June 2014.

\bibitem{LIU20091921}
Shaobo Liu, Lizhong Yang, Tingyong Fang, and Jian Li.
\newblock Evacuation from a classroom considering the occupant density around
  exits.
\newblock {\em Physica A: Statistical Mechanics and its Applications},
  388(9):1921 -- 1928, 2009.

\bibitem{MA20102101}
Jian Ma, Wei guo Song, Jun Zhang, Siu ming Lo, and Guang xuan Liao.
\newblock k-nearest-neighbor interaction induced self-organized pedestrian
  counter flow.
\newblock {\em Physica A: Statistical Mechanics and its Applications},
  389(10):2101 -- 2117, 2010.

\bibitem{MATLAB}
{MATLAB}.
\newblock \verb|https://www.mathworks.com/products/matlab.htm|.

\bibitem{measles_terminal1}
{Notes from the Field: Measles Transmission at a Domestic Terminal Gate in an
  International Airport -- United States, January 2014}.
\newblock \verb|https://www.cdc.gov/mmwr/preview/mmwrhtml/mm6350a9.htm|.

\bibitem{measles_terminal2}
{Notes from the Field: Measles Transmission in an International Airport at a
  Domestic Terminal Gate -- April-May 2014}.
\newblock \verb|https://www.cdc.gov/mmwr/preview/mmwrhtml/mm6424a6.htm|.

\bibitem{Moussad2755}
Mehdi Moussa{\"\i}d, Dirk Helbing, Simon Garnier, Anders Johansson, Maud Combe,
  and Guy Theraulaz.
\newblock Experimental study of the behavioural mechanisms underlying
  self-organization in human crowds.
\newblock {\em Proceedings of the Royal Society of London B: Biological
  Sciences}, 276(1668):2755--2762, 2009.

\bibitem{Mubayi2010}
Anuj Mubayi, Christopher~Kribs Zaleta, Maia Martcheva, and Carlos
  Castillo-Chavez.
\newblock A cost-based comparison of quarantine strategies for new emerging
  diseases.
\newblock {\em Mathematical Biosciences and Engineering},
  7(1551-0018-2010-3-687):687, 2010.

\bibitem{Muller2000}
Johannes Mueller, Mirjam Kretzschmar, and Klaus Dietz.
\newblock Contact tracing in stochastic and deterministic epidemic models.
\newblock {\em Mathematical Biosciences}, 164:39--64, 2000.

\bibitem{Parisi2007343}
D.R. Parisi and C.O. Dorso.
\newblock Morphological and dynamical aspects of the room evacuation process.
\newblock {\em Physica A: Statistical Mechanics and its Applications},
  385(1):343 -- 355, 2007.

\bibitem{T-Krithika}
K.~Rathinakumar.
\newblock {\em A Microscopic Approach to Pedestrian Dynamics and the Onset of
  Disease Spreading}.
\newblock PhD thesis, University of Houston, 2019.

\bibitem{Rivers2014}
C.M. Rivers, E.T. Lofgren, M.~Marathe, S.~Eubank, and B.L. Lewis.
\newblock Modeling the impact of interventions on an epidemic of ebola in
  {Sierra Leone and Liberia}.
\newblock {\em PLoS Curr.,6}, 2014.

\bibitem{Rohani2000}
P~Rohani, D.J.D Earn, and B.T Grenfell.
\newblock The impact of immunisation on pertussis transmission in england and
  wales.
\newblock {\em Lancet}, 355:285–286, 2000.

\bibitem{Schadschneider2011}
Andreas Schadschneider, Wolfram Klingsch, Hubert Kl{\"u}pfel, Tobias Kretz,
  Christian Rogsch, and Armin Seyfried.
\newblock {\em Evacuation Dynamics: Empirical Results, Modeling and
  Applications}, pages 517--550.
\newblock Springer New York, New York, NY, 2011.

\bibitem{Seyfried2009}
Armin Seyfried, Oliver Passon, Bernhard Steffen, Maik Boltes, Tobias Rupprecht,
  and Wolfram Klingsch.
\newblock New insights into pedestrian flow through bottlenecks.
\newblock {\em Transportation Science}, 43(3):395--406, 2009.

\bibitem{5773492}
A.~Shende, M.~P. Singh, and P.~Kachroo.
\newblock Optimization-based feedback control for pedestrian evacuation from an
  exit corridor.
\newblock {\em IEEE Transactions on Intelligent Transportation Systems},
  12(4):1167--1176, Dec 2011.

\bibitem{TAJIMA2002709}
Yusuke Tajima, Kouhei Takimoto, and Takashi Nagatani.
\newblock Pattern formation and jamming transition in pedestrian counter flow.
\newblock {\em Physica A: Statistical Mechanics and its Applications},
  313(3):709 -- 723, 2002.

\bibitem{TurnerPenn}
Alasdair Turner and Alan Penn.
\newblock Encoding natural movement as an agent-based system: An investigation
  into human pedestrian behaviour in the built environment.
\newblock {\em Environment and Planning B: Planning and Design},
  29(4):473--490, 2002.

\bibitem{Ward150703}
Jonathan~A. Ward, Andrew~J. Evans, and Nicolas~S. Malleson.
\newblock Dynamic calibration of agent-based models using data assimilation.
\newblock {\em Royal Society Open Science}, 3(4), 2016.

\bibitem{5339199}
S.~Xu and H.~B.~L. Duh.
\newblock A simulation of bonding effects and their impacts on pedestrian
  dynamics.
\newblock {\em IEEE Transactions on Intelligent Transportation Systems},
  11(1):153--161, March 2010.

\bibitem{Yu2005}
W.~J. Yu, R.~Chen, L.~Y. Dong, and S.~Q. Dai.
\newblock Centrifugal force model for pedestrian dynamics.
\newblock {\em Phys. Rev. E}, 72:026112, Aug 2005.

\bibitem{Zhang2011}
J.~Zhang, W.~Klingsch, A.~Schadschneider, and A.~Seyfried.
\newblock Transitions in pedestrian fundamental diagrams of straight corridors
  and {T-junctions}.
\newblock {\em J. Stat. Mech.}, 2011:P06004, 2011.

\bibitem{6248013}
B.~Zhou, X.~Wang, and X.~Tang.
\newblock Understanding collective crowd behaviors: Learning a mixture model of
  dynamic pedestrian-agents.
\newblock In {\em 2012 IEEE Conference on Computer Vision and Pattern
  Recognition}, pages 2871--2878, June 2012.

\end{thebibliography}
\bibliographystyle{plain}

\end{document}